\pdfoutput=1
\documentclass[english,pre,a4paper,aps,reprint]{revtex4-2}
\usepackage[T1]{fontenc}
\usepackage{amsmath}
\usepackage{amssymb}
\usepackage{graphicx}

\makeatletter

\usepackage{hyperref}

\makeatother

\usepackage[english]{babel}
\begin{document}
\title{Memory effects in a sequence of measurements of non-commuting observables}
\author{Sophia M. Walls and Ian J. Ford}
\affiliation{Department of Physics and Astronomy, University College London, Gower
Street, London, WC1E 6BT, United Kingdom}
\begin{abstract}
We use continuous, stochastic quantum trajectories within a framework
of quantum state diffusion (QSD) to describe alternating measurements
of two non-commuting observables. Projective measurement of an observable
completely destroys memory of the outcome of a previous measurement
of the conjugate observable. In contrast, measurement under QSD is
not projective and it is possible to vary the rate at which information
about previous measurement outcomes is lost by changing the strength
of measurement. We apply our methods to a spin 1/2 system and a spin
1 system undergoing alternating measurements of the $S_{z}$ and $S_{x}$
spin observables. Performing strong $S_{z}$ measurements and weak
$S_{x}$ measurements on the spin 1 system, we demonstrate return
to the same eigenstate of $S_{z}$ to a degree beyond that expected
from projective measurements and the Born rule. Such a memory effect
appears to be greater for return to the $\pm1$ eigenstates than the
$0$ eigenstate. Furthermore, the spin 1 system follows a measurement
cascade process where an initial superposition of the three eigenstates
of the observable evolves into a superposition of just two, before
finally collapsing into a single eigenstate, giving rise to a distinctive
pattern of evolution of the spin components.
\end{abstract}
\maketitle

\section{Introduction}

\begin{figure}
\begin{centering}
\includegraphics[width=1\columnwidth]{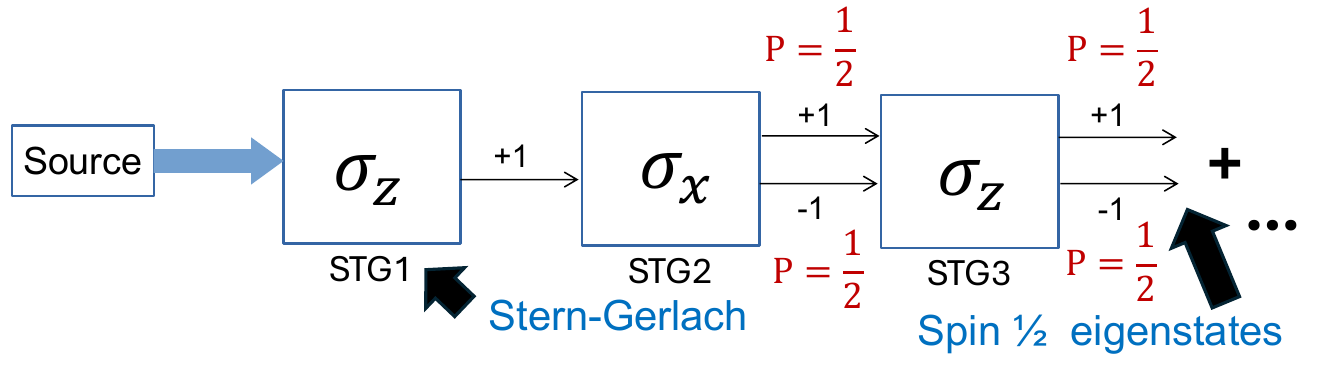}
\par\end{centering}
\caption{A demonstration of memory erasure in a sequence of projective measurements
of non-commuting observables $\sigma_{z}$ and $\sigma_{x}$. Stern-Gerlach
devices 1 (STG1) and 3 (STG3) conduct $\sigma_{z}$ measurements whilst
STG2 performs a $\sigma_{x}$ measurement. The system is measured
to be in the +1 state at STG1 but in a second $\sigma_{z}$ measurement
at STG3 the system has equal probability of being in the +1 or $-1$
state. The cause of this `forgetting' is the projective $\sigma_{x}$
measurement at STG2. \label{fig:non_comm_obs}}
\end{figure}

Non-commuting or conjugate observables are considered to be a hallmark
of quantum mechanics. Measurement of one is conventionally taken to
erase any information about its conjugate partner \citep{norsen_2017,nielsen2002a}.
Figure \ref{fig:non_comm_obs} demonstrates how a $\sigma_{z}$ measurement
outcome at Stern-Gerlach device 1 (STG1) acting on a beam of spin
1/2 particles is `forgotten' because of the measurement of non-commuting
observable $\sigma_{x}$ at STG2. The system is measured at STG1 to
be in the $+1$ eigenstate of $\sigma_{z}$ yet the second $\sigma_{z}$
measurement at STG3 has an equal likelihood of $\pm1$ outcomes. Naturally
this is reflected in the mathematics by the fact that the $|+\rangle_{x}$
and $|-\rangle_{x}$ spin 1/2 eigenstates can be written as an equal
amplitude superposition of $|+\rangle_{z}$ and $|-\rangle_{z}$ eigenstates:
$|+\rangle_{x}=\frac{1}{\sqrt{2}}(|+\rangle_{z}+|-\rangle_{z})$,
where the suffixes indicate the measured observable.

Whilst conventional quantum mechanics focuses on projective, instantaneous
measurements, and average outcomes which mask what happens during
single measurements, more generalised descriptions of measurement
exist as well as a wider scope of measurement types \citep{nielsen2002a}.
Weak measurements are a frequently cited example \citep{pan.etal_2023,rozema.etal_2012,malik.etal_2014,lundeen.etal_2011,lundeen.steinberg_2009,lundeen.bamber_2012,hosten.kwiat_2008,piacentini.etal_2016}.
Whilst projective measurements extract maximum information about a
system, weak measurements extract partial information, causing less
disturbance to the system itself. The system collapses over a finite
time frame or may not completely collapse at an eigenstate at all. 

Weak measurements relax many quantum foundational principles related
to conjugate variables. For example Calder\'{o}n-Losada \textit{et
al} were able to assign simultaneously measured values of two non-commuting
observables to particles in an EPR (Einstein-Podolsky-Rosen) experiment
using weak measurements \citep{calderon-losada.etal_2020}. Rozema
\textit{et al }demonstrated a violation of Heisenberg's uncertainty
principle using weak measurement \citep{rozema.etal_2012}. Similarly,
weak measurements have been used to establish `weak value' correlators
across measurement outcomes or to demonstrate violations of the Leggett-Garg
or Bell inequalities \citep{chantasri.etal_2018,atalaya.etal_2018,jordan.buttiker_2005,suzuki.etal_2016,piacentini.etal_2016b,suzuki.etal_2012,dressel2011,goggin2011,williams2008,palacios-laloy.etal_2010}.
Non-commuting observables have also been studied in the context of
testing quantumness, quantum randomness using the Kochen-Spekker theorem
and the relation between the EPR settings and Heisenberg's uncertainty
principle \citep{kulikov.etal_2017,simon.etal_2000,huang2003a,howell2004,garcia-pintos2016,sindelka.moiseyev_2018}.
Hossenfelder \textit{et al }proposed a scheme where a sequence of
measurements of non-commuting observables might be used to examine
correlations in measurement outcomes of the same observable at different
times \citep{hossenfelder_2011}.

In this paper we study the destruction of prior information in a sequence
of alternating measurements of non-commuting observables. We consider
how such information loss may be reduced if weak measurements are
used instead of strong, projective ones.

In order to investigate such a scenario, a framework where measurement
occurs over a finite amount of time is required, one where single
evolution pathways for a system may be tracked \citep{minev.etal_2019,jordan_2013,murch.etal_2013}.
Many techniques exist for generating quantum trajectories, such as
the Feynman-Vernon path integral formalism, hierarchical equations
of motion, and various unravellings of the Lindblad master equation
\citep{gambetta.etal_2008,gneiting.etal_2021,plenio.knight_1998,tanimura_2015,lindblad1976a,devega2017,feynman1963,li2022b,wiseman1993a}.
The Liouville-von Neumann equation is a further example, though it
does not preserve positivity and thus the trajectories generated cannot
be considered as physically realistic \citep{teixeira.etal_2021,chantasri.jordan_2015,gneiting.etal_2021,plenio.knight_1998,tanimura_2015,lindblad1976a}.

In this work we generate stochastic, continuous and diffusive quantum
trajectories using quantum state diffusion (QSD) \citep{percival1998a}.
Quantum state diffusion is a framework which describes measurement
as a continuous and gradual process. Similarly to Gisin and Percival,
we interpret measurement in this framework to be a consequence of
a system-environment interaction, where we consider the environment
to act as a (classical) measurement apparatus \citep{percival1998a,gisin1992b}.
This is to be contrasted with Gambetta, Wiseman and Jacobs who consider
the system's state to be conditioned on the measured state of the
environment which then contains information about the state of the
system \citep{gambetta.etal_2008,wiseman1993a,wiseman1996a,jacobs2014}.

The interaction with the environment drives the system towards an
eigenstate of the observable in a continuous and diffusive manner,
such that the completion of measurement takes a finite amount of time
\citep{walls.etal_2024,gisin1997}. System evolution is described
by a stochastic differential equation (SDE) that includes noise terms
representing its interactions with the environment and hence the effect
of measurement. In the absence of interactions with the environment,
evolution is described by the (deterministic) von Neumann equation.
The SDE generates single stochastic trajectories and the average over
all trajectories generated by the environmental noise yields the Lindblad
master equation \citep{gisin1997}. Many different unravellings of
the Lindblad master equation exist, corresponding to different measurement
schemes such as direct photon detection, heterodyne or homodyne measurements
and more recently, the jump-time unravelling. QSD is an example of
a continuous unravelling of the Lindblad master equation.

The possibility of simultaneous measurement of conjugate variables
through weak measurement has proved to have a useful application in
direct measurement of the elements of a density matrix of a quantum
system and quantum state tomography \citep{thekkadath.etal_2016,ber.etal_2013,brodutch.cohen_2017,kim.etal_2018,lundeen.etal_2011,clarke.ford_2023},
as well as quantum error correction and quantum cybersecurity \citep{atalaya.etal_2017,berta.etal_2010,bera.etal_2017,prevedel.etal_2011}
which rely on simultaneous or alternating measurements of non-commuting
observables. Quantum cybersecurity algorithms and cryptography which
rely on the principle that measurements of conjugate observables destroy
information about one another, could be left vulnerable to attack
by the possibility of incomplete measurement \citep{hacohen-gourgy.etal_2016,jordan.buttiker_2005,piacentini.etal_2016,hofmann_2014,brodutch.cohen_2017,rozema.etal_2012,chen.etal_2019,mitchison.etal_2007,ochoa.etal_2018,ozawa_2003,ruskov.etal_2010,sindelka.moiseyev_2018,suzuki.etal_2016}.

We employ stochastic quantum trajectories to investigate memory effects
in a series of alternate measurements of two non-commuting observables.
We perform strong measurements of the first observable and weak measurements
of the second. Incomplete collapse of the system to an eigenstate
of the second observable could preserve information about the first
observable. For example, in Fig. \ref{fig:non_comm_obs}, if the measurement
at STG2 were weak, the system might return more often than expected
at STG3 to the same eigenstate of $\sigma_{z}$ it was in at STG1.
Memory effects in incomplete measurements have been studied before
using quantum trajectories in relation to the partial quantum Zeno
effect, for example \citep{peres.ron_1990,zhang.etal_2019}, and there
have been investigations of the reversal of quantum measurements whereby
the system returns to a previous state after a partial collapse \citep{katz.etal_2008,korotkov.jordan_2006,murch.etal_2013,kim.etal_2012}.
Such memory effects can be observed only when studying single system
evolution pathways as opposed to average system behaviour, and with
continuous and gradual wavefunction collapse instead of instantaneous
and discontinuous evolution.

We study two separate systems: a spin 1/2 system and a spin 1 system,
both undergoing alternating measurements of spin component $S_{z}$
and $S_{x}$ observables. We consider cases of equal measurement strength
of both observables, and more particularly a situation involving strong
$S_{z}$ measurements and a varying strength of $S_{x}$ measurement.
We explore system return to the same eigenstate of $S_{z}$, given
a particular strength of measurement of $S_{x}$, and the pattern
of system exploration of its parameter space.

The plan for the paper is as follows: section \ref{sec:Quantum-State-Diffusion}
outlines the QSD framework and general SDE used to generate quantum
trajectories; section \ref{sec:System-Specification} introduces the
SDEs specific to the spin 1/2 system and the spin 1 system as well
as describing the protocol for generating sequential measurements
of non-commuting observables; and finally sections \ref{sec:Results}
and \ref{sec:Conclusion} present our results and conclusions.

\section{Quantum State Diffusion and Stochastic Lindblad Equations \label{sec:Quantum-State-Diffusion}}

Measurement in QSD arises from an interaction between an open quantum
system and a classical environment, the effect of which is to evolve
the system continuously and stochastically towards one of the eigenstates
of the measured observable. Our interpretation is that the uncertain
initial state of the environment, together with complex interaction
dynamics, is responsible for the stochastic evolution of the system
state, rather like the stochastic evolution of an open classical system.
In such circumstances the open quantum system state can be modelled
using a randomly evolving (reduced) density operator, or matrix, $\rho$.
We describe the mean (Markovian) evolution of $\rho$ in a time increment
$dt$ using a superoperator $\hat{S}[\rho]$ defined in terms of Kraus
operators $M_{j}$ \citep{tong2006a,nielsen2002a}:

\begin{align}
\hat{S}[\rho(t)] & =\sum_{j}M_{j}(dt)\rho(t)M_{j}^{\dagger}(dt)\nonumber \\
 & =\sum_{j}p_{j}\frac{M_{j}(dt)\rho(t)M_{j}^{\dagger}(dt)}{Tr(M_{j}(dt)\rho(t)M_{j}^{\dagger}(dt))},\label{eq:kraus_operator_sum-1}
\end{align}
where we have written $p_{j}=Tr(M_{j}\rho M_{j}^{\dagger})$. Such
an evolution is positive and trace-preserving when the Kraus operators
satisfy the completeness condition $\sum_{j}M_{j}^{\dagger}M_{j}=\mathbb{I}$.
The evolution is Markovian since the future state of the system depends
only on its current state $\rho(t)$.

 Equation (\ref{eq:kraus_operator_sum-1}) can be interpreted as
an average over transitions of the current state of the system $\rho(t)$
to one of several possible states $M_{j}\rho(t)M_{j}^{\dagger}/Tr(M_{j}\rho(t)M_{j}^{\dagger})$
in the time interval $dt$, each taking place with probability $p_{j}=Tr(M_{j}\rho(t)M_{j}^{\dagger})$
\citep{walls.etal_2024}. To generate diffusive, continuous trajectories
we employ Kraus operators that differ incrementally from (proportionality
to) the identity:

\begin{equation}
M_{j}\equiv M_{k\pm}=\frac{1}{\sqrt{2}}(\mathbb{I}+A_{k\pm}),\label{eq:1+A}
\end{equation}
where
\begin{equation}
A_{k\pm}=-iH_{s}dt-\frac{1}{2}L_{k}^{\dagger}L_{k}dt\pm L_{k}\sqrt{dt}.\label{eq:Ak}
\end{equation}
The index $k$ specifies Lindblad channels, each of which is associated
with two Kraus operators labelled by $\pm$ \citep{walls.etal_2024,matos.etal_2022,gross2018a,clarke2022a,jacobs2014}.
The system Hamiltonian is denoted by $H_{s}$ and the Lindblad operators
by $L_{k}$. 

In QSD we consider an ensemble of states that the system could adopt,
with the average density matrix over the ensemble then denoted as
$\bar{\rho}(t)$. A single member of the ensemble evolves in a timestep
according to the action of a single Kraus operator $M_{j}$, representing
the transition

\begin{equation}
\rho(t+dt)=\rho(t)+d\rho=\frac{M_{j}\rho(t)M_{j}^{\dagger}}{Tr(M_{j}\rho(t)M_{j}^{\dagger})},\label{eq:single_kraus-1}
\end{equation}
and a sequence of transitions then yields a stochastic trajectory.
Note that such a map with a Kraus operator of the form given in Eqs.
(\ref{eq:1+A}) and (\ref{eq:Ak}) has been demonstrated to preserve
positivity \citep{walls.etal_2024}. The average evolution over all
possible system transitions may then be shown to satisfy the Lindblad
master equation \citep{matos.etal_2022}

\begin{equation}
d\bar{\rho}=-i[H_{s},\bar{\rho}]dt+\sum_{k}\left(L_{k}\bar{\rho}L_{k}^{\dagger}-\frac{1}{2}\{L_{k}^{\dagger}L_{k},\bar{\rho}\}\right)dt,
\end{equation}
while a single trajectory is a solution to the following SDE for quantum
state diffusion:

\begin{align}
d\rho & =-i[H_{s},\rho]dt+\sum_{k}\Big((L_{k}\rho L_{k}^{\dagger}-\frac{1}{2}\{L_{k}^{\dagger}L_{k},\rho\})dt\nonumber \\
 & \qquad+\left(\rho L_{k}^{\dagger}+L_{k}\rho-Tr[\rho(L_{k}+L_{k}^{\dagger})]\rho\right)dW_{k}\Big).\label{eq:SME-1}
\end{align}
The evolution of $\rho$ is thus an It\^{o} process with Wiener increments
$dW_{k}$, one for each Lindblad operator or channel $k$ \citep{gisin1997,percival1998a},
and arising from system-environment interactions.

It is well known that the quantum operation in Eq. (\ref{eq:kraus_operator_sum-1})
is invariant under a unitary transformation of the Kraus operators:
$M_{j}\to M_{j}^{\prime}=\sum_{m}U_{jm}M_{m}$. Thus the Kraus operator
representation in Eqs. (\ref{eq:1+A}) and (\ref{eq:Ak}) would not
appear to be unique. For example, the same Lindblad equation for the
evolution of the mean density matrix $\bar{\rho}(t)$ would emerge
for Kraus operator pairs: $M_{k\pm}^{\prime}=\frac{1}{\sqrt{2}}\left(M_{k+}\pm M_{k-}\right)$
or more explicitly $M_{k+}^{\prime}=\mathbb{I}-iH_{s}dt-\frac{1}{2}L_{k}^{\dagger}L_{k}dt$
and $M_{k-}^{\prime}=L_{k}\sqrt{dt}$.

However, what is unique about the choice of $M_{k\pm}$ is that they
generate continuous trajectories: they both are proportional to the
identity when $dt\to0$, while $M_{k-}^{\prime}$, in contrast, generates
jumps. Since we take the view that individual trajectories of the
density matrix must be continuous, arising from a stochastic Lindblad
equation that is diffusive, unitary freedom in the choice of Kraus
operators is not available here.

\section{System Specification and Dynamics \label{sec:System-Specification}}

For simplicity we consider a system with $H_{s}=0$ and two Lindblad
operators $L_{z}=aS_{z}$ and $L_{x}=bS_{x}$ where the $S_{i}$ are
spin component operators and $a(t)$ and $b(t)$ are scalar, time-dependent
coefficients representing system-environment couplings. Each interaction
is associated with Wiener increments $dW_{z}$ and $dW_{x}$, respectively,
in the dynamics. When inserted into Eq. (\ref{eq:SME-1}), we obtain
the following SDE \citep{clarke2022a}

\begin{align}
d\rho= & a^{2}\left(S_{z}\rho S_{z}-\frac{1}{2}\rho S_{z}^{2}-\frac{1}{2}S_{z}^{2}\rho\right)dt+\nonumber \\
+ & b^{2}\left(S_{x}\rho S_{x}-\frac{1}{2}\rho S_{x}^{2}-\frac{1}{2}S_{x}^{2}\rho\right)dt+\nonumber \\
+ & a\left(\rho S_{z}+S_{z}\rho-2\langle S_{z}\rangle\rho\right)dW_{z}+\nonumber \\
+ & b\left(\rho S_{x}+S_{x}\rho-2\langle S_{x}\rangle)\rho\right)dW_{x},\label{eq:two_opSDE}
\end{align}
where $\langle S_{i}\rangle=Tr\left(\rho S_{i}\right)$. Note that
we do not consider $\langle S_{i}\rangle$ to be an `expectation value'
in the conventional sense: namely as a statistical property of the
dynamics. Rather we regard it as a stochastically evolving physical
attribute of the quantum state of the system \citep{walls.etal_2024}.

\begin{figure}
\centering{}\includegraphics[width=1\columnwidth]{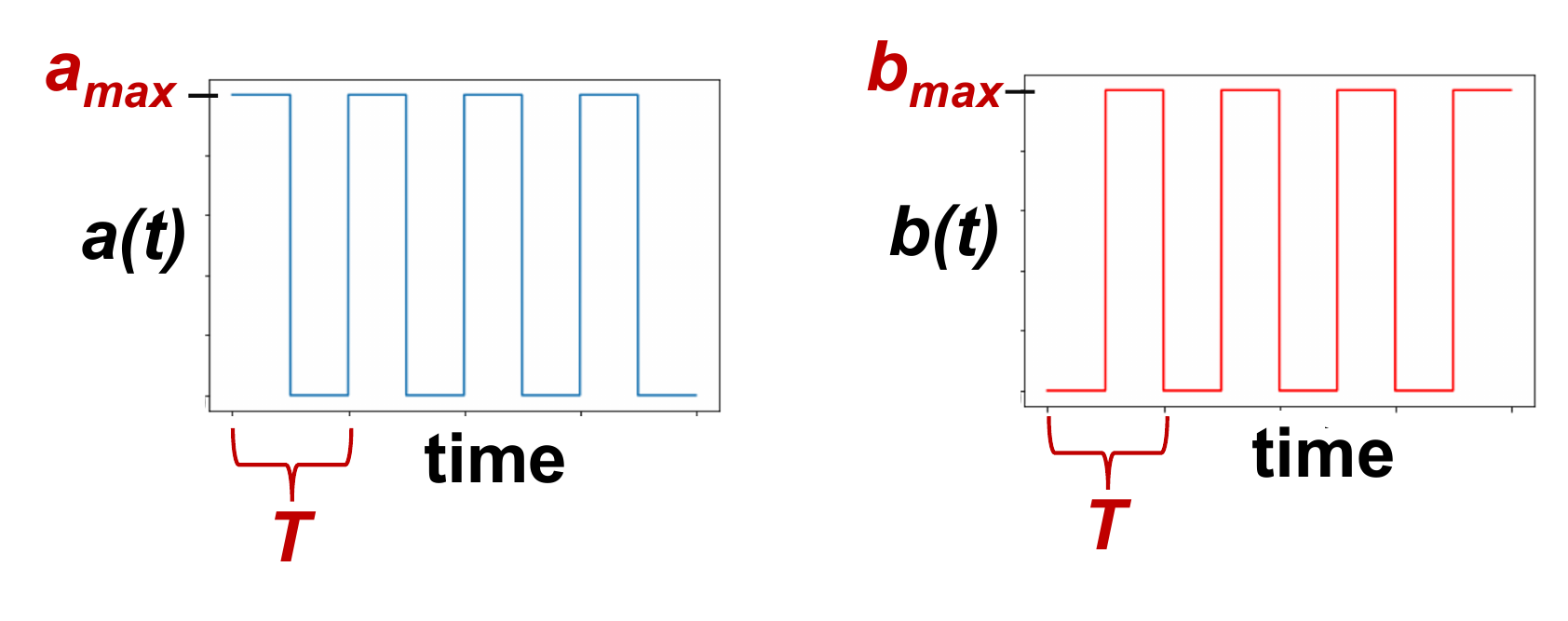}\caption{System-environment couplings $a(t)$ and $b(t)$ for the observables
$S_{z}$ and $S_{x}$, respectively. In order to produce alternating
measurements of the non-commuting observables, they are modelled as
box functions such that when $a(t)=1,\,b(t)=0$ and vice-versa. The
measurement strengths are defined by the maximum values of $a(t)$
and $b(t)$ $(a_{max}$ and $b_{max}$) and the period of the box
function $T$, as given in Eq. (\ref{eq:mx}). \label{fig:box_fns}}
\end{figure}

In order to generate a sequence of alternating measurements of the
non-commuting observables $S_{z}$ and $S_{x}$ we take the system-environment
couplings $a(t)$ and $b(t)$ to be time-varying box functions subject
to $a(t)\ne0,b(t)=0$ and vice-versa (see Figure \ref{fig:box_fns}).
The period of the box function $T$ in conjunction with $a_{max}$
(the maximum value of $a(t)$) defines the $S_{z}$ measurement strength
$M_{z}$, whilst $b_{max}$ (the maximum value of $b(t)$) defines
the $S_{x}$ measurement strength $M_{x}$:

\begin{align}
M_{z}=a_{max}^{2}\frac{T}{2}\quad, & \quad M_{x}=b_{max}^{2}\frac{T}{2}.\label{eq:mx}
\end{align}
The period of measurement of one of the observables is half of the
box function period.

\subsection{Spin 1/2 system}

The density matrix of a spin 1/2 system can be written $\rho=\frac{1}{2}(\mathbb{I}+\boldsymbol{r}\cdot\boldsymbol{\sigma})$
where $\boldsymbol{r}=(r_{x},r_{y},r_{z})$ is the coherence vector
$\boldsymbol{r}=Tr(\rho\boldsymbol{\sigma})$ and $\boldsymbol{\sigma}=(\sigma_{x},\sigma_{y},\sigma_{z})$
is a vector of Pauli matrices. Using $d\boldsymbol{r}=Tr(d\rho\boldsymbol{\sigma})$,
and Eq. (\ref{eq:two_opSDE}), we can derive SDEs for the coherence
vector components. We choose $r_{y}=0,\,dr_{y}=0$ for simplicity
and obtain the following SDEs, using $S_{z}=\frac{1}{2}\sigma_{z}$
and $S_{x}=\frac{1}{2}\sigma_{x}$ in Eq. (\ref{eq:two_opSDE}) \citep{clarke.ford_2023}:

\begin{alignat}{1}
dr_{z}= & 2a\left(1-r_{z}^{2}\right)dW_{z}-2b^{2}r_{z}dt-2br_{z}r_{x}dW_{x},\nonumber \\
dr_{x}= & 2b\left(1-r_{x}^{2}\right)dW_{x}-2a^{2}r_{x}dt-2ar_{x}r_{z}dW_{z}.
\end{alignat}
Stochastic trajectories in terms of the coherence vector components
$\left(r_{x},r_{z}\right)$ are then generated using these SDEs. Note
that $r_{x},r_{z}$ define the density matrix at each point in time
since $r_{y}=0.$

\subsection{Spin 1 system\label{subsec:Spin-1-system}}

The generalised Bloch sphere formalism is used to represent the density
matrix of a spin 1 system. This permits any $3\times3$ density matrix
to be written in terms of the Gell-Mann matrices $\lambda_{i}$ through

\begin{equation}
\rho=\frac{1}{3}(\mathbb{I}+\sqrt{3}\boldsymbol{R}\cdot\boldsymbol{\lambda}),\label{eq:rho_spin_1}
\end{equation}
where $\boldsymbol{R}=(s,m,u,v,k,x,y,z)$ is an eight dimensional
coherence (or Bloch) vector and the Gell-Mann matrices form the elements
of the vector $\boldsymbol{\lambda}=(\lambda_{1},\lambda_{2},\lambda_{3},\lambda_{4},\lambda_{5},\lambda_{6},\lambda_{7},\lambda_{8})$
(see Appendix \ref{subsec:Appendix-A} for details). SDEs for the
evolution of the coherence vector components can then be constructed
and are given in Appendix \ref{subsec:Appendix-A}.

The $S_{x}$ and $S_{z}$ operators for a spin 1 system each have
three eigenstates: $|-1\rangle_{x},|0\rangle_{x},|1\rangle_{x}$ and
$|-1\rangle_{z},|0\rangle_{z},|1\rangle_{z}$, respectively. The spin
components $\langle S_{x}\rangle$, $\langle S_{y}\rangle$ and $\langle S_{z}\rangle$
can be written:
\begin{align}
\langle S_{x}\rangle & =\sqrt{\frac{2}{3}}(x+s)\nonumber \\
\langle S_{y}\rangle & =\sqrt{\frac{2}{3}}(m+y)\nonumber \\
\langle S_{z}\rangle & =\frac{u}{\sqrt{3}}+z.\label{eq:spin 1 spin components}
\end{align}

We take the initial state of the system to be an equal mixture of
the three eigenstates of $S_{z}$: $|\rho_{{\rm start}}\rangle=\frac{1}{3}(|1\rangle_{z}\langle1|_{z}+|0\rangle_{z}\langle0|_{z}+|-1\rangle_{z}\langle-1|_{z})$.
As a result, $\langle S_{y}\rangle$ remains zero throughout, as well
as the variables $k,m$ and $y$, leaving five variables as active
participants in the dynamics. Stochastic quantum trajectories are
then generated using the SDEs for these variables. Rather than imagining
their evolution in the five dimensional space of $(s,u,v,x,z)$, we
project them into the $(\langle S_{x}\rangle,\langle S_{z}\rangle)$
space which offers a more insightful understanding of the dynamics.
Note that the eigenstates $|0\rangle_{x}$ and $|0\rangle_{z}$ both
lie at $\langle S_{x}\rangle=0,\langle S_{z}\rangle=0$, but are orthogonal.
Indeed $\langle S_{x}\rangle=0,\langle S_{z}\rangle=0$ can describe
many different states, including the initial state of the system.

\section{Dwell times and cumulative probability densities \label{sec:Results}}

We wish to understand how alternating measurements of conjugate observables
destroy information about one another. As illustrated in Figure \ref{fig:non_comm_obs},
conventional projective measurement at STG2 implies that the information
gathered at STG1 is `forgotten' such that measurement of the same
observable at STG3 yields an uncorrelated outcome. Weak measurements
might not cause the system to collapse completely to an eigenstate
of $S_{x}$ since information about the system is being extracted
at a slower rate. As a result, subsequent $S_{z}$ measurement might
be biassed towards a return to its previous state. 

\begin{figure}
\begin{centering}
\includegraphics[width=1\columnwidth]{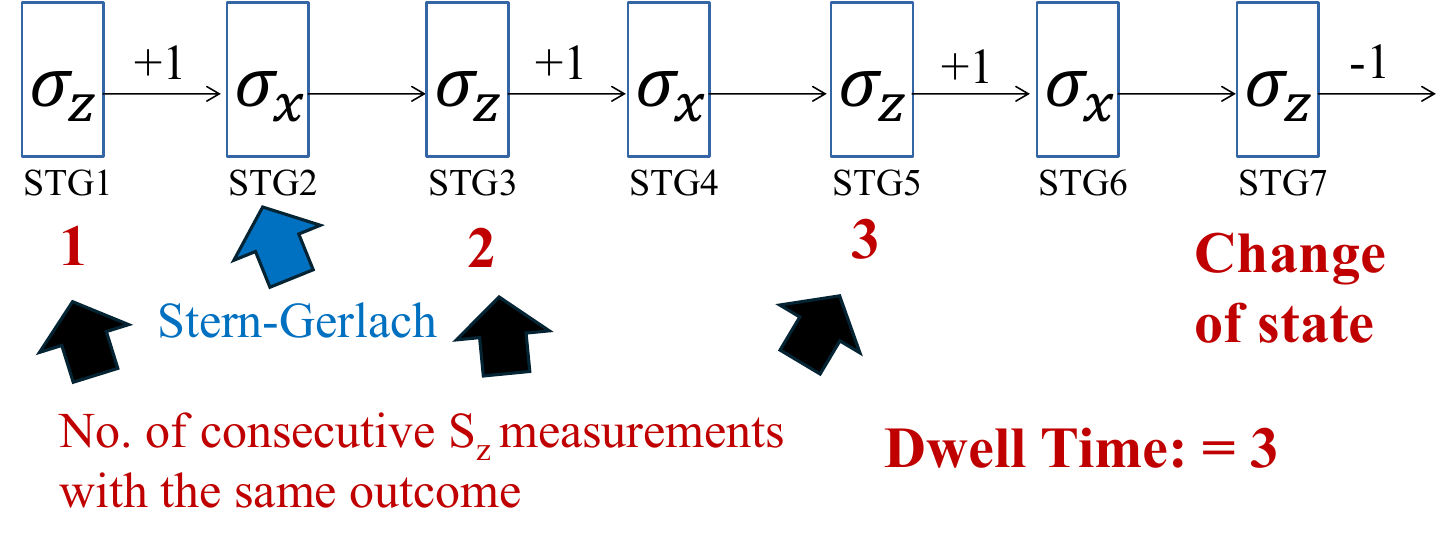}
\par\end{centering}
\caption{Investigating the number of repeated $S_{z}$ measurement outcomes
before a spin 1/2 system is able to adopt a different $S_{z}$ eigenstate.
\label{fig:return_eigen}}
\end{figure}

From the quantum trajectories, we produce a time series of measurement
outcomes of $S_{z}$. The time parameter is defined by the count of
$S_{z}$ measurements that have taken place in the sequence of alternating
measurements of $S_{x}$ and $S_{z}$. For each of the eigenstates
of $S_{z}$, we can calculate the number of consecutive returns to
the same eigenstate and we shall refer to this as the \textit{dwell
time}. We calculate how long the system \textit{dwells} at a particular
eigenstate in spite of attempts to measure a conjugate observable.
Figure \ref{fig:return_eigen} illustrates an example: at STG7 the
system is found to take the $-1$ eigenstate, and therefore the dwell
time in the +1 eigenstate since its initiation at STG1, is 3. We investigate
this with a high strength of measurement of $S_{z}$ while varying
the strength of the $S_{x}$ measurements.

We perform dwell time calculations using strong $S_{z}$ measurements
and a range of strengths of $S_{x}$ measurements. In this situation,
there is ample time for the system to approach very close to an eigenstate
of $S_{z}$ for every measurement, therefore the \textit{dwell time}
does not depend on a precise definition of eigenstate adoption. The
dwell time is expected to depend on the strength of measurement of
$S_{x}$.

We also use trajectories to generate cumulative probability densities
in the $(\langle S_{x}\rangle,\langle S_{z}\rangle)$ plane to illustrate
the behaviour of the system. We do this for equal strength of measurement
of the two observables. 

\subsection{Spin 1/2 system}

\begin{figure}
\begin{centering}
\includegraphics[width=1\columnwidth]{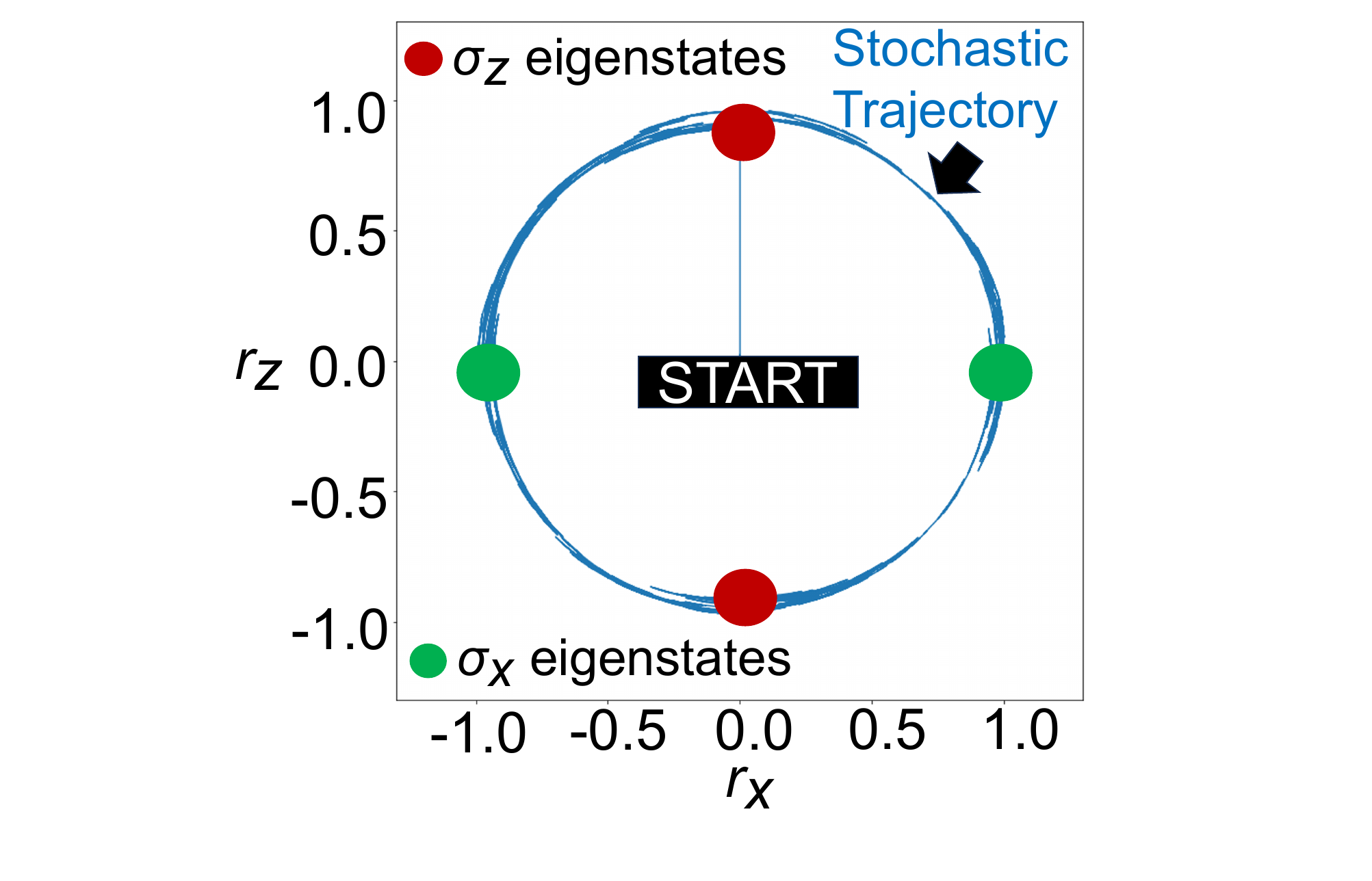}
\par\end{centering}
\caption{An example stochastic trajectory (in blue) of the coherence vector
components $(r_{x},r_{z})$ for the spin 1/2 system, starting in the
mixed state at the centre of the Bloch sphere. When a measurement
is performed, the system is purified and travels towards the circumference,
where it remains. \label{fig:rx_rz_example_traj}}
\end{figure}

Numerical trajectories of $r_{x}$ and $r_{z}$ were generated for
the spin 1/2 system undergoing sequential, alternating measurements
of $\sigma_{x}$ and $\sigma_{z}$. We begin with a case where the
measurement strengths of $\sigma_{x}$ and $\sigma_{z}$ are equal.
An example trajectory is shown in Figure \ref{fig:rx_rz_example_traj}.
The system begins at the centre of the circle in the totally mixed
state. An initial $\sigma_{z}$ measurement is performed with equal
probability of arriving at either $r_{z}=1$ or $r_{z}=-1$. As $\sigma_{z}$
and $\sigma_{x}$ measurements are performed, the system purifies
and evolves towards the circumference of the circle, where it remains.
The $\sigma_{x}$ eigenstates lie at the equator, equidistant from
the $\sigma_{z}$ eigenstates at the poles. Starting from one of the
$\sigma_{x}$ eigenstates there is an equal chance for the system
to travel towards either of the $\sigma_{z}$ eigenstates during an
$\sigma_{z}$ measurement (and vice-versa).

When the measurement strength is weak, the measurement process may
be incomplete: the system may not arrive at the eigenstate of the
observable that is being measured. Figures \ref{fig:ex_traj_spin_half}a
and b illustrate a $\sigma_{z}$ measurement of the system, between
times 18 and 19, that causes the system to evolve towards the $r_{z}=+1$
eigenstate while $r_{x}$ tends towards zero. A $\sigma_{x}$ measurement
is then performed between times 19 and 20, leading the system away
from $r_{x}=0$, but it does not unequivocally arrive at an eigenstate
of $\sigma_{x}$. Within the time interval 20-21, a second $\sigma_{z}$
measurement is performed and the system moves back to the $r_{z}=+1$
eigenstate it adopted when the first $\sigma_{z}$ measurement was
conducted during the time interval 18-19. A $\sigma_{x}$ measurement
is made once again between times 21 and 22, causing the system to
travel towards a $\sigma_{x}$ eigenstate but again it is not successful
in reaching it. As a result, when a third $\sigma_{z}$ measurement
is performed in interval 22-23, the system again returns to the same
eigenstate as in the previous two measurements of $\sigma_{z}$. Since
the $\sigma_{x}$ measurements fail to take the system to a $\sigma_{x}$
eigenstate, the system is allowed to return to the same eigenstate
of $\sigma_{z}$ multiple times.

This can also be understood using Figure \ref{fig:rx_rz_example_traj}.
Let us take the system to be in the vicinity of the north pole after
a $\sigma_{z}$ measurement. A subsequent $\sigma_{x}$ measurement
might not bring the system to an eigenstate in the time available,
so that the system remains in the top half of the circle. This means
that when a second $\sigma_{z}$ measurement is conducted, the system
is more likely to return to the $r_{z}=+1$ eigenstate than the $r_{z}=-1$
eigenstate, since it is closer. This is how memory might emerge.
\begin{center}
\begin{figure}
\begin{centering}
\includegraphics[width=1\columnwidth]{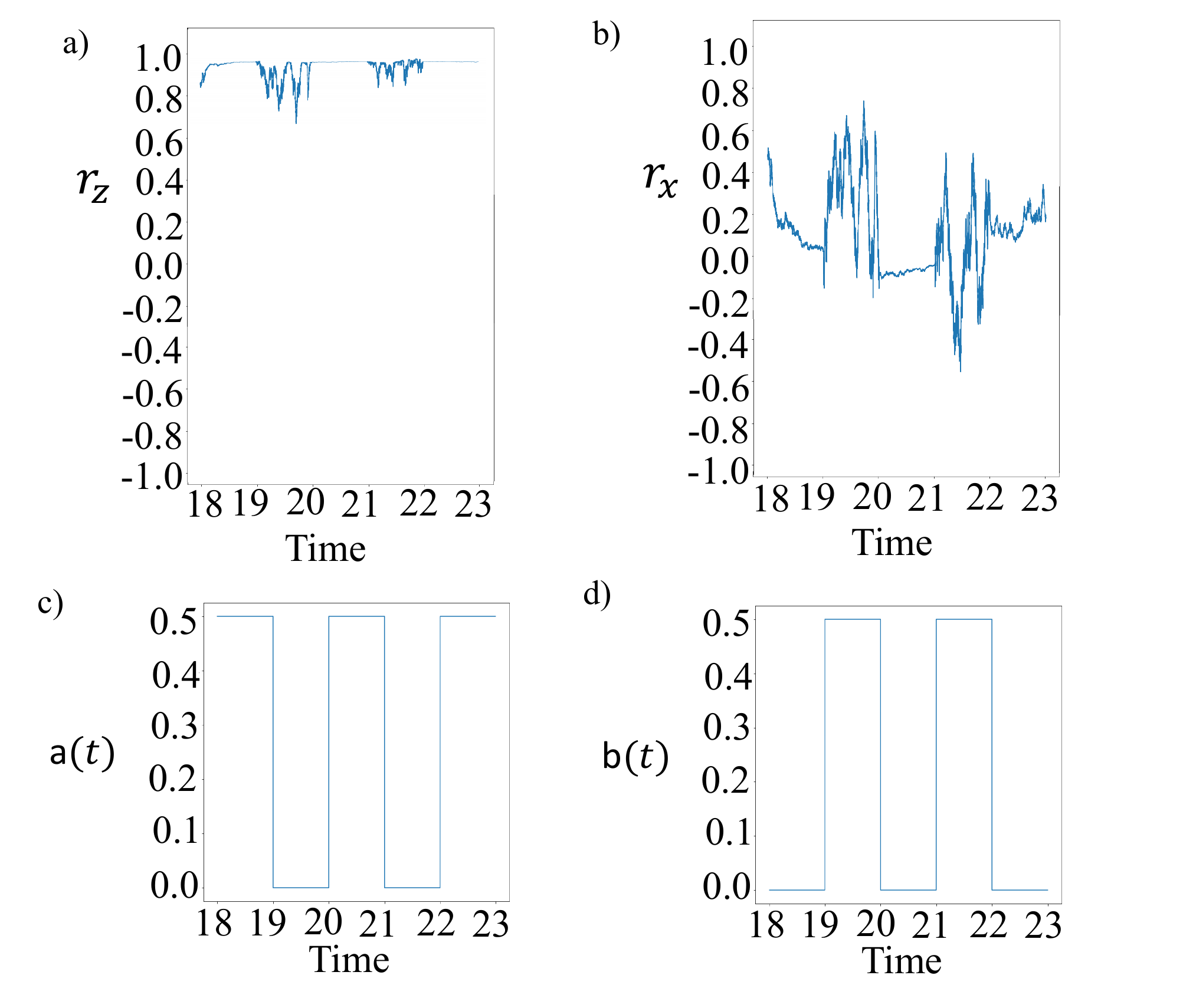}
\par\end{centering}
\caption{a) and b) An example trajectory illustrating how the coherence vector
components $r_{z}$ and $r_{x}$ vary with time. c) evolution of the
coupling $a(t)$ and d) the coupling $b(t)$. Measurement strengths:
$M_{x}=M_{z}=\frac{1}{4}$ and time-step 0.0001 were used. A measurement
of $\sigma_{z}$ starts at times 18, 20 and 22, whilst a measurement
of $\sigma_{x}$ starts at 19, 21 and 23. \label{fig:ex_traj_spin_half}}
\end{figure}
\par\end{center}

Cumulative probability densities over the ($r_{x},r_{z})$ plane are
illustrated in Figure \ref{fig:spin_half_prob_density} for $M_{x}=M_{z}$.
As the measurement strengths  are increased, the system spends more
time in the vicinity of the eigenstates of $\sigma_{z}$ and $\sigma_{x}$
and the time spent in each of their respective eigenstates is equal
(Figures \ref{fig:spin_half_prob_density} a and b). As the measurement
strengths are reduced, the system explores the circumference of the
circle more uniformly (Figure \ref{fig:spin_half_prob_density} d)
or shows a bias towards certain eigenstates of the observables than
others. For example, Figure \ref{fig:spin_half_prob_density}c shows
a tendency to spend more time at the $r_{x}=+1$ eigenstate of $\sigma_{x}$
and the $r_{z}=-1$ eigenstate of $\sigma_{z}$.
\begin{center}
\begin{figure}
\begin{centering}
\includegraphics[width=1\columnwidth]{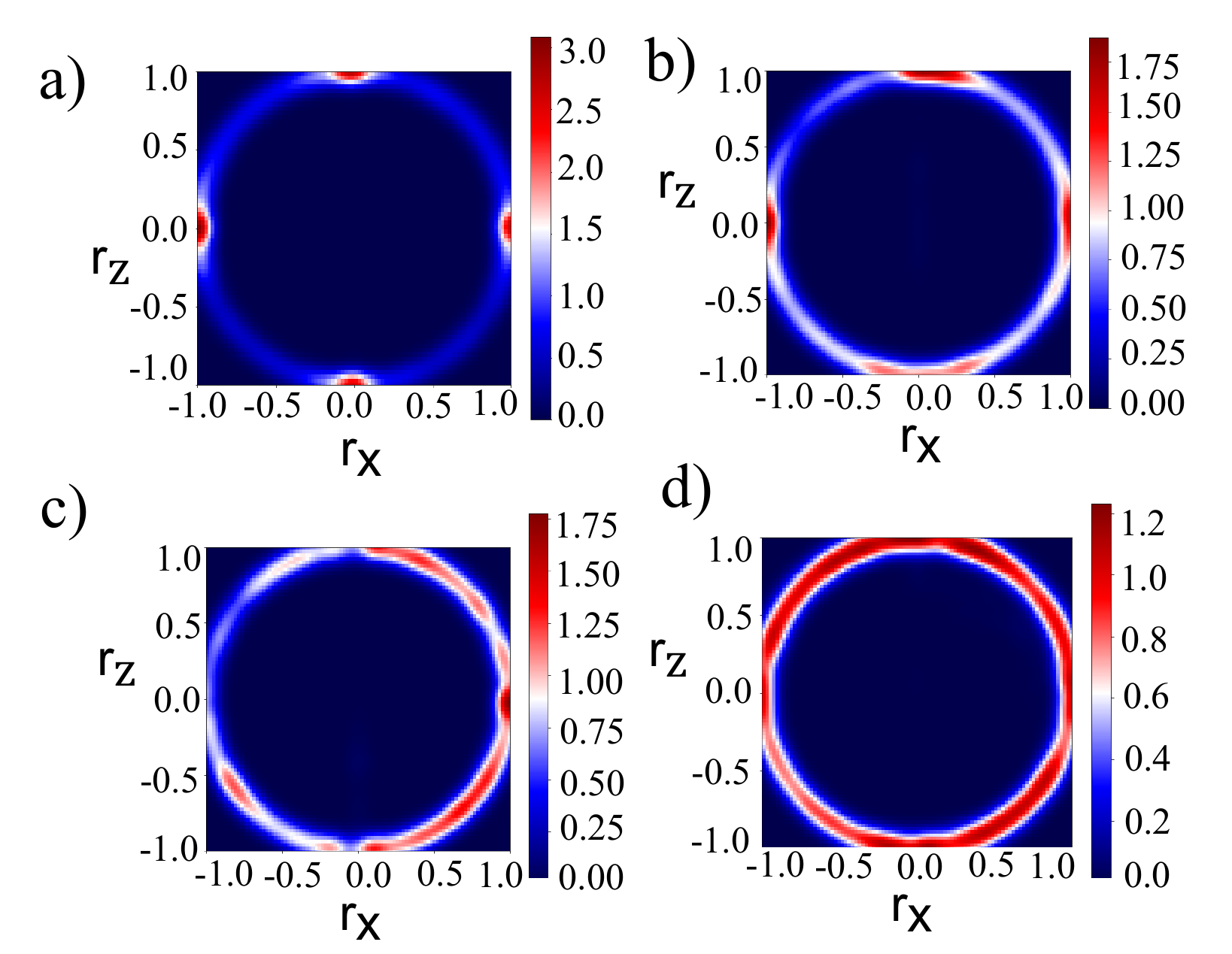}
\par\end{centering}
\caption{Color. Cumulative probability densities in the $(r_{x},r_{z})$ phase
space, arising from continuous stochastic trajectories for the spin
1/2 system under sequential measurement. Measurement strengths $M_{x}=M_{z}$
are as follows: a) 1, b) 0.5, c) 0.25, d) 0.1875. Duration 200, time-step
0.0001. \label{fig:spin_half_prob_density}}
\end{figure}
\par\end{center}

\begin{figure}
\begin{centering}
\includegraphics[width=1\columnwidth]{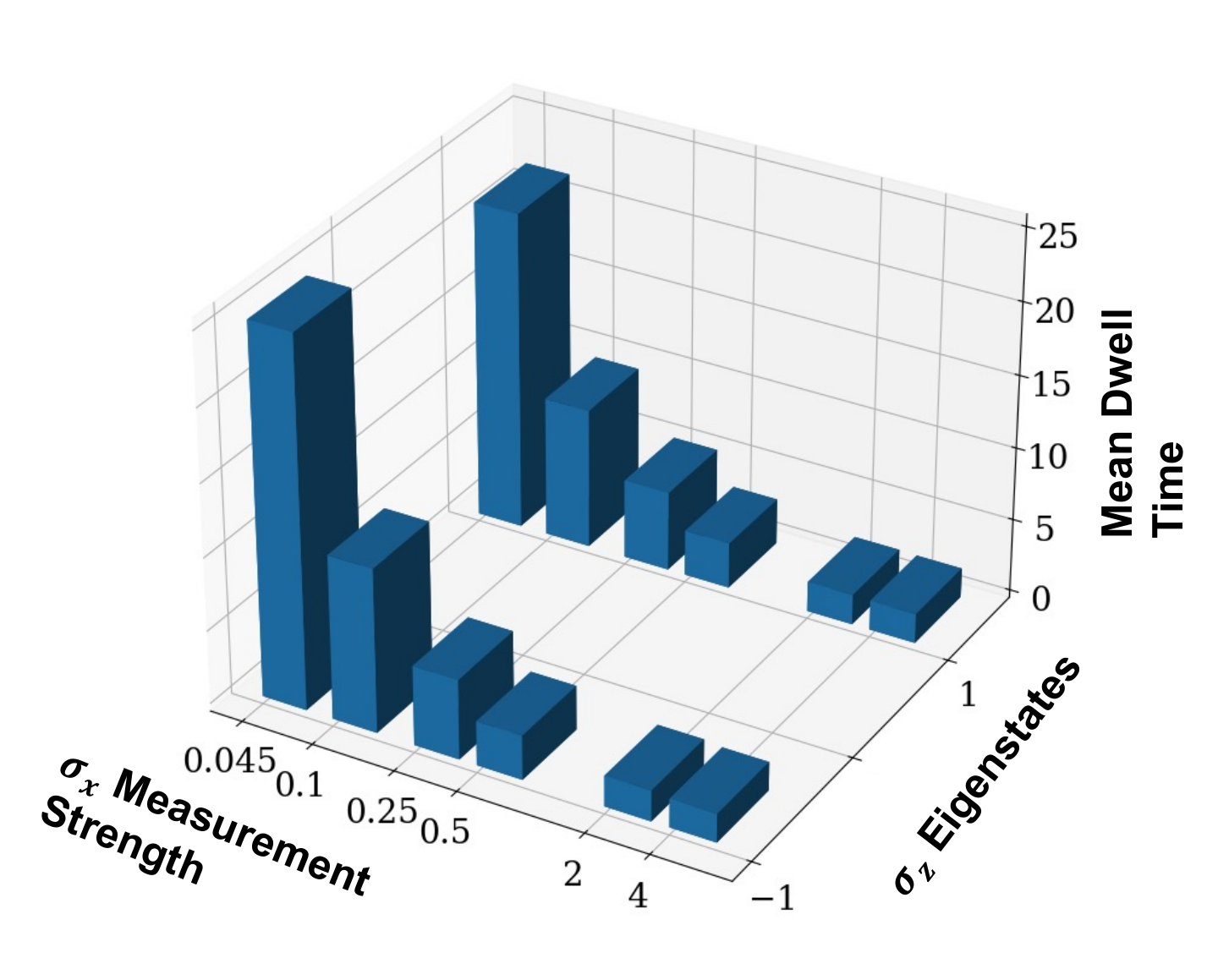}
\par\end{centering}
\caption{The mean dwell time for the spin 1/2 system for the two eigenstates
of $\sigma_{z}$ and for a range of $\sigma_{x}$ measurement strengths.
The dwell time is the number of consecutive $\sigma_{z}$ measurements
for which the system is able to return to the same eigenstate. 5000
$\sigma_{z}$ measurements were used to calculate the mean. The $\sigma_{z}$
measurement strength was 32.\label{fig:spin_half_dwell_time}}
\end{figure}

The stabilisation of the system at a particular eigenstate of $\sigma_{z}$,
illustrated in Figure \ref{fig:ex_traj_spin_half}, is now investigated
by employing strong $\sigma_{z}$ measurements and weak $\sigma_{x}$
measurements in the sequence of alternating measurements. We calculate
the mean dwell time: the average number of times that the system is
able to return to the same eigenstate of $\sigma_{z}$. Figure \ref{fig:spin_half_dwell_time}
illustrates the mean dwell time for the spin 1/2 system for both eigenstates
of $\sigma_{z}$, and for varying $\sigma_{x}$ measurement strength.

From Figure \ref{fig:spin_half_dwell_time} it can be seen that the
mean dwell time for both eigenstates increases as the $\sigma_{x}$
measurement strength is decreased whilst keeping the $\sigma_{z}$
measurement strength high, indicating that the system is more biassed
towards return to the same eigenstate of $\sigma_{z}$ for longer
measurement sequences. The dwell time is roughly equal for both eigenstates
for all $\sigma_{x}$ measurement strengths.

\subsection{Spin 1 system}

\begin{figure}
\begin{centering}
\includegraphics[width=1\columnwidth]{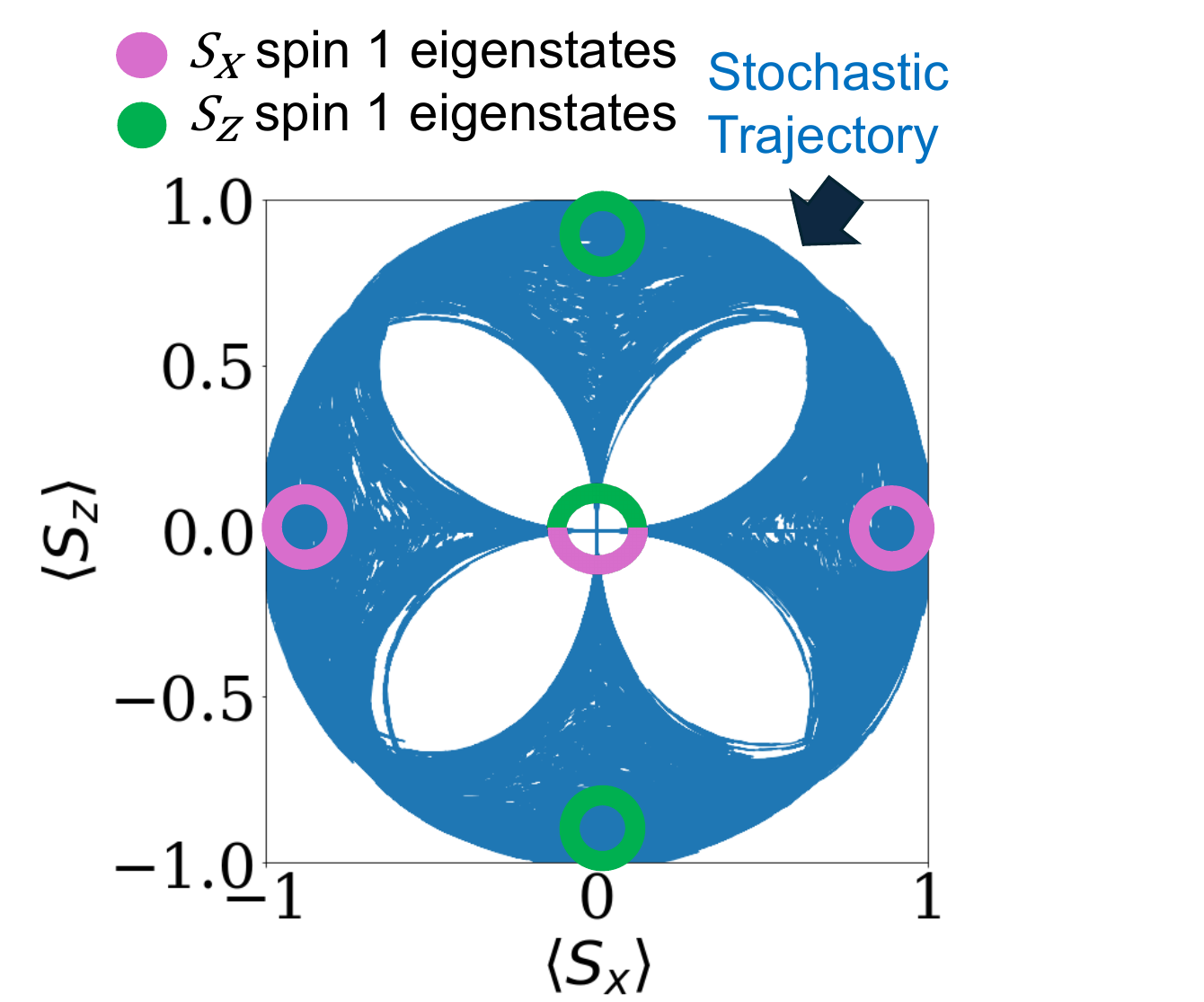}
\par\end{centering}
\caption{An example trajectory of the spin 1 system, starting in the mixed
state $\rho_{\text{start}}=\frac{1}{3}(|1\rangle_{z}\langle1|_{z}+|0\rangle_{z}\langle0|_{z}+|-1\rangle_{z}\langle-1|_{z}))$
illustrated in terms of the spin components $(\langle S_{x}\rangle,\langle S_{z}\rangle)$.
Measurement strength: $M_{x}=M_{z}=8$, duration 500, time-step 0.0001.
The $S_{x}$ eigenstates are illustrated in pink and the $S_{z}$
eigenstates in green. The $|\pm1\rangle_{x}$ eigenstates lie at the
equator and the $|\pm1\rangle_{z}$ eigenstates at the poles. The
$|0\rangle_{z}$ and the $|0\rangle_{x}$ eigenstates are both at
the centre but are mutually orthogonal. \label{fig:spin_1_eigenstates}}
\end{figure}

Numerical trajectories for the parameters of the spin 1 density matrix
were used to compute trajectories for the spin components $(\langle S_{x}\rangle,\langle S_{z}\rangle)$.
Figure \ref{fig:spin_1_eigenstates} depicts an example trajectory
with the $S_{x}$ eigenstates illustrated in pink: the $|\pm1\rangle_{x}$
eigenstates at the equator and the $|0\rangle_{x}$ eigenstate at
the centre. The $S_{z}$ eigenstates are shown in green: the $|\pm1\rangle_{z}$
eigenstates at the poles and the $|0\rangle_{z}$ eigenstate also
at the centre, noting that it is orthogonal to $|0\rangle_{x}$.

\begin{figure}
\begin{centering}
\includegraphics[width=1\columnwidth]{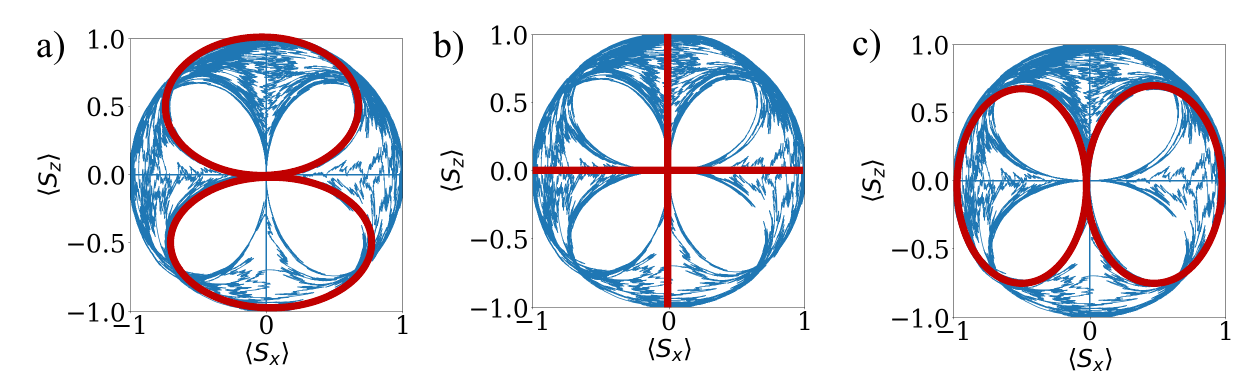}
\par\end{centering}
\caption{Color. Example trajectory of the spin 1 system in the space of spin
components $(\langle S_{x}\rangle,\langle S_{z}\rangle)$, with measurement
strengths: $M_{x}=M_{z}=8$, duration 80, time-step 0.0001. a) The
red ellipse in the upper section is the locus of a superposition of
$|+1\rangle_{z}$ and $|0\rangle_{z}$ eigenstates, and the lower
red ellipse a superposition of $|-1\rangle_{z}$ and $|0\rangle_{z}$.
b) The vertical red line represents superposition between $|-1\rangle_{z}$
and $|1\rangle_{z}$ eigenstates and the horizontal line a superposition
between $|-1\rangle_{x}$ and $|1\rangle_{x}$. c) The red ellipse
on the left is a locus of a superposition between $|-1\rangle_{x}$
and $|0\rangle_{x}$ eigenstates, and the red ellipse on the right
is a superposition between $|1\rangle_{x}$ and $|0\rangle_{x}$.
\label{fig:ent_traj_ex}}
\end{figure}

The system is initiated at the centre of the circle, in the mixed
state $\rho_{\mathrm{start}}=\frac{1}{3}(|1\rangle_{z}\langle1|_{z}+|0\rangle_{z}\langle0|_{z}+|-1\rangle_{z}\langle-1|_{z}))$
such that there is an equal probability of reaching any of the $S_{z}$
eigenstates following an $S_{z}$ measurement. The trajectory shows
an intriguing avoidance of certain regions, giving rise to a petal-like
pattern of exploration. We find that the pattern is influenced by
combinations of two of the three eigenstates of $S_{x}$ and of $S_{z}$.
For example, in Figure \ref{fig:ent_traj_ex}a a superposition of
the $|+1\rangle_{z}$ and $|0\rangle_{z}$ eigenstates is represented
by the red ellipse in the upper section and a superposition of the
$|-1\rangle_{z}$ and $|0\rangle_{z}$ eigenstates by the red ellipse
in the lower section. The horizontal straight line in Figure \ref{fig:ent_traj_ex}b
illustrates a superposition between the $|1\rangle_{x}$ and $|-1\rangle_{x}$
eigenstates and the vertical line a superposition between the $|1\rangle_{z}$
and $|-1\rangle_{z}$ eigenstates.

Similarly, in Figure \ref{fig:ent_traj_ex}c a superposition of the
$|+1\rangle_{x}$ and $|0\rangle_{x}$ eigenstates is illustrated
by the red ellipse in the right hand section and a superposition of
the $|-1\rangle_{x}$ and $|0\rangle_{x}$ eigenstates by the red
ellipse on the left.

The boundaries of the avoided petal-like regions correspond to combinations
of the red ellipses in Figures \ref{fig:ent_traj_ex}a and \ref{fig:ent_traj_ex}c.
We argue that this pattern stems from the way in which the spin 1
system collapses to an eigenstate.

\begin{figure}
\begin{centering}
\includegraphics[width=1\columnwidth]{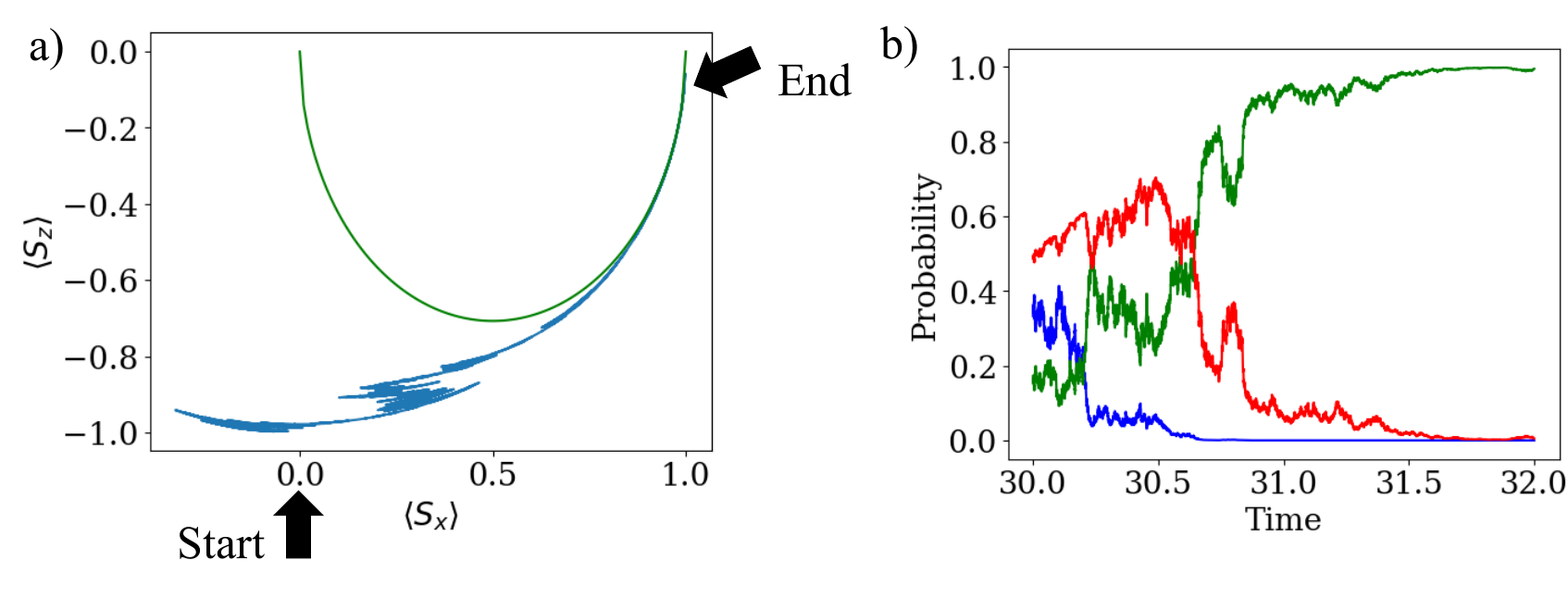}
\par\end{centering}
\caption{a) An example of a collapse cascade in the spin 1 system. The system
begins at the $|-1\rangle_{z}$ eigenstate and then undergoes a $S_{x}$
measurement. The green semi-elliptical locus represents a superposition
between the $|0\rangle_{x}$ and $|1\rangle_{x}$ eigenstates. During
the $S_{x}$ measurement, the system moves first to a point on this
locus, as the probability of the $|-1\rangle_{x}$ eigenstate decreases
to zero. The system then remains on the green locus until collapsing
at the $|1\rangle_{x}$ eigenstate. b) The evolving probabilities
of the $|1\rangle_{x}$, $|0\rangle_{x}$ and $|-1\rangle_{x}$ eigenstates
of $S_{x}$ in green, red and blue, respectively. $S_{x}$ measurement
strength 2. \label{fig:mes_cascade}}
\end{figure}

To understand this, consider a situation where the system takes the
$|-1\rangle_{z}$ eigenstate after undergoing a measurement of $S_{z}$,
establishing a starting point for a subsequent measurement of $S_{x}$,
an example pathway for which is depicted in Figure \ref{fig:mes_cascade}a.
From the $|-1\rangle_{z}$ eigenstate, an $S_{x}$ measurement leads
to probabilities of $\frac{1}{4}$, $\frac{1}{2}$ and $\frac{1}{4}$
for collapsing to the $|1\rangle_{x}$, $|0\rangle_{x}$ and $|-1\rangle_{x}$
eigenstates, respectively. Figure \ref{fig:mes_cascade}b illustrates
how the probabilities of collapse to the eigenstates of $S_{x}$ evolve
over the course of the measurement. Initially the probabilities are
indeed those defined by the $|-1\rangle_{z}$ eigenstate expressed
in the $S_{x}$ basis. The system evolves in this case by first eliminating
the probability of collapse to $|-1\rangle_{x}$, taking the system
to a superposition between the $|1\rangle_{x}$ and $|0\rangle_{x}$
eigenstates alone, depicted in Figures \ref{fig:ent_traj_ex}c as
a red ellipse on the right and in Figure \ref{fig:mes_cascade}a as
a green semi-elliptical locus. The system remains in a superposition
of the two eigenstates, and confined to the green locus, until collapsing
to only one of them, which in Figure \ref{fig:mes_cascade}a happens
to be the $|1\rangle_{x}$ eigenstate. Figure \ref{fig:mes_cascade}b
reflects this final collapse by showing the probabilities of the $|1\rangle_{x}$
and the $|0\rangle_{x}$ eigenstates rising to one and falling to
zero, respectively.

We could think of this as a `collapse cascade': a multistep process
of wavefunction collapse. The system collapses first onto a subspace
spanned by two of the eigenstates of the observable, before ultimately
completely collapsing at one eigenstate. The petal regions in Figure
\ref{fig:spin_1_eigenstates} are avoided because once the system
has arrived at a point on one of the red ellipses depicted in Figure
\ref{fig:ent_traj_ex}, typically coming from the region outside the
petal-like regions, then a move within the petals would mean re-introducing
a non-zero amplitude for the third eigenstate of the observable. This
would be contrary to the nature of the measurement dynamics. Indeed,
when in a two eigenstate superposition, the drift and diffusion coefficients
along the direction of the eigenstate with zero amplitude are both
zero. Thus eigenstates in the superposition are typically eliminated
one by one.

Note that if the system had started in the $|0\rangle_{z}$ eigenstate,
then it would already be in a superposition of just $|1\rangle_{x}$
and $|-1\rangle_{x}$ since $|0\rangle_{z}=-\frac{1}{\sqrt{2}}|1\rangle_{x}+\frac{1}{\sqrt{2}}|-1\rangle_{x}$.
The system would then evolve along the horizontal line depicted in
Figure \ref{fig:ent_traj_ex}b, leading towards one or other of $|\pm1\rangle_{x}$.
This situation therefore skips step one of the measurement collapse
cascade. Entering the petal-like regions would still represent a regression
in the measurement collapse process through introducing a non-zero
amplitude for the $|0\rangle_{x}$ eigenstate, and is incompatible
with the dynamics. 

\begin{figure}
\begin{centering}
\includegraphics[width=1\columnwidth]{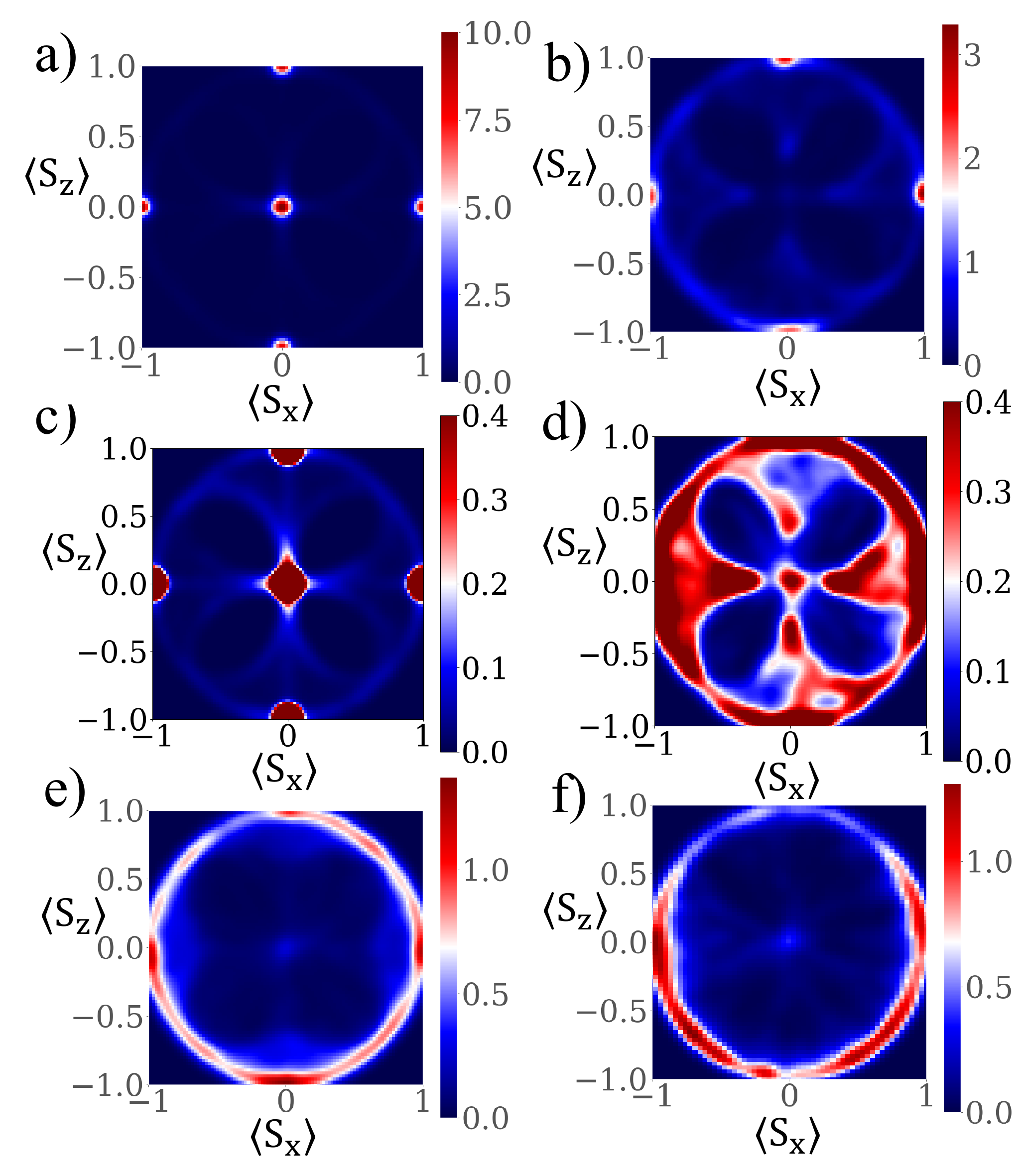}
\par\end{centering}
\caption{Color. Trajectories of the spin 1 system in terms of the spin components
$(\langle S_{x}\rangle,\langle S_{z}\rangle)$ are used to generate
a cumulative probability density. Duration 500, time-step 0.0001.
Measurement strengths for both observables: a) 8, b) 2, c) 8, showing
cumulative probability density with a range of 0-0.4 d) 2, similarly
with a range of probability density of 0-0.4 e) 1, f) 0.5. \label{fig:spin_1_density_plots}}
\end{figure}

As with the spin 1/2 system, we can use a long trajectory to compute
a cumulative probability density that characterises the evolution
of the spin 1 system in the space of the spin components $(\langle S_{x}\rangle,\langle S_{z}\rangle)$.
Figure \ref{fig:spin_1_density_plots} depicts how the cumulative
probability density varies with measurement strength. For measurement
strengths of 8 for both observables, shown in Figure \ref{fig:spin_1_density_plots}a,
the highest probability density is found around the eigenstates of
$S_{x}$ and $S_{z}$. In accordance with the Born rule, the probability
around the centre is twice as large as around the other eigenstates,
since this is the location of both the $|0\rangle_{x}$ and the $|0\rangle_{z}$
eigenstates. As the measurement strength is decreased, the probability
density spreads away from the eigenstates such as in Figures \ref{fig:spin_1_density_plots}e
and f. It is notable, in particular, that the |$0\rangle_{x}$ and
the $|0\rangle_{z}$ eigenstates are visited significantly less frequently
than the $|\pm1\rangle_{x}$ and $|\pm1\rangle_{z}$ eigenstates when
the measurement strength is reduced: the cumulative probability at
the centre of the circle is low in comparison to the other eigenstates.
It would therefore seem that at low measurement strength there is
less adherence to the Born rule.

Figures \ref{fig:spin_1_density_plots}c and d are the same as a and
b, respectively, but are displayed with a maximum cumulative probability
density of 0.4 for the purpose of examining the paths followed by
the system in greater detail. The avoided petal-like regions are clear
in Figures \ref{fig:spin_1_density_plots}b,c,d,e. For high measurement
strength, illustrated in detail in Figure \ref{fig:spin_1_density_plots}c,
the cumulative probability density along the edges of the petals increases
towards the $\langle S_{x}\rangle=\langle S_{z}\rangle=0$ region
and is higher than in the vicinities of the $\pm1$ eigenstates of
the two observables. However, for a lower measurement strength, depicted
in Figures \ref{fig:spin_1_density_plots}b and d, the cumulative
probability density spreads further away from the centre, creating
a distinct region of exclusion around $\langle S_{x}\rangle=\langle S_{z}\rangle=0$.
For even lower measurement strength, shown in Figure \ref{fig:spin_1_density_plots}e,
the cumulative probability density is pushed further towards the circumference
of the circle. This suggests that when measurement strength is low,
the system tends to follow pathways that lead to collapse at the eigenstates
of $S_{x}$ and $S_{z}$ that lie on the circumference of the circle
rather than at the centre. At the lowest measurement strength, in
Figure \ref{fig:spin_1_density_plots}f, the petal patterns are hardly
visible, suggesting that the two-step measurement cascade process
is not so rigidly followed. In Figure \ref{fig:spin_1_density_plots}f
the system appears to collapse less often at the $|1\rangle_{z}$
and $|0\rangle_{z}$ eigenstates than at $|-1\rangle_{z}$, though
this could be a result of insufficient sampling. 

\begin{figure}
\begin{centering}
\includegraphics[width=1\columnwidth]{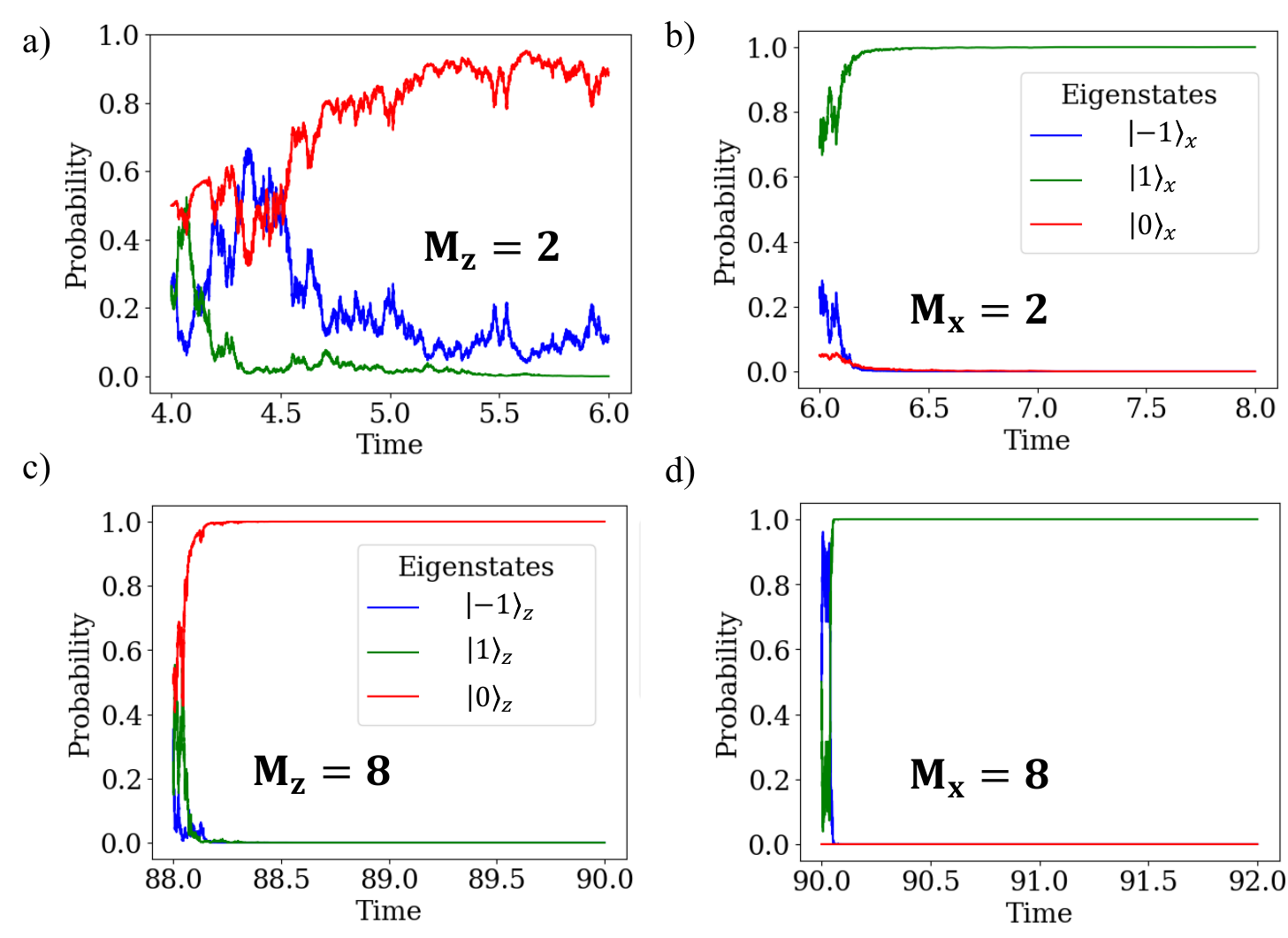}
\par\end{centering}
\caption{a) Collapse probabilities to the $S_{z}$ eigenstates during an $S_{z}$
measurement with measurement strength $M_{z}=2$. b) Collapse probabilities
of $S_{x}$ eigenstates during an $S_{x}$ measurement with measurement
strength $M_{x}=2$ that takes place directly after the evolution
shown in a). c) As a) with measurement strength $M_{z}=8$. d) As
b) with measurement strength $M_{x}=8$. Plots a) and c) share a legend,
as do plots b) and d). Note that the selection of $|0\rangle_{z}$
in a) and c) takes longer than the selection of $|1\rangle_{x}$ in
b) and d) \label{fig:zero_eig_avoided}}
\end{figure}

We can speculate about the reason for the lower frequency of visits
to the $|0\rangle_{x}$ and $|0\rangle_{z}$ eigenstates during weak
measurement compared with the strong measurement case. At low measurement
strength, the system might not always be able to collapse at an eigenstate
over the time-frame of the measurement. For weak measurement strength,
Figure \ref{fig:zero_eig_avoided}a illustrates collapse probabilities
to the $S_{z}$ eigenstates over the course of an $S_{z}$ measurement
and Figure \ref{fig:zero_eig_avoided}b illustrates the collapse probabilities
to the $S_{x}$ eigenstates in a $S_{x}$ measurement performed directly
afterwards. The system reaches the vicinity of $|0\rangle_{z}$ after
an $S_{z}$ measurement. Since $|0\rangle_{z}$ written in the $S_{x}$
basis $\left(|0\rangle_{z}=-\frac{1}{\sqrt{2}}|1\rangle_{x}+\frac{1}{\sqrt{2}}|-1\rangle_{x}\right)$
suggests equal probabilities of collapse to $|1\rangle_{x}$ and $|-1\rangle_{x}$,
the system then has a probability of collapse to $|0\rangle_{x}$
which is close to zero during the subsequent $S_{x}$ measurement,
depicted in Figure \ref{fig:zero_eig_avoided}b. It therefore takes
little time for the system to reach a state of superposition of just
the $|1\rangle_{x}$ and $|-1\rangle_{x}$ eigenstates. In effect,
the system is given a `head start' in its collapse to the $|1\rangle_{x}$
and $|-1\rangle_{x}$ eigenstates since the probability of the $|0\rangle_{x}$
eigenstate at the beginning of the $S_{z}$ measurement was already
very close to zero.

In contrast, if the system lies initially in the vicinity of a $|1\rangle_{x}$
or $|-1\rangle_{x}$ eigenstate, the subsequent $S_{z}$ measurement
is a two-step process since probabilities of all three $S_{z}$ eigenstates
are substantial, for example $|-1\rangle_{x}=\frac{1}{2}|1\rangle_{z}+\frac{1}{\sqrt{2}}|0\rangle_{z}+\frac{1}{2}|-1\rangle_{z}$.
The system has to dedicate a significant amount of time to the first
stage of the measurement cascade, the arrival at a point on one of
the two-eigenstate superposition loci. If the $|0\rangle_{z}$ eigenstate
is accessible largely when the system starts an $S_{z}$ measurement
from $|\pm1\rangle_{x}$ rather than $|0\rangle_{x}$ then its selection
is always subject to this overhead, in contrast to the selection of
$|\pm1\rangle_{x}$ starting from $|0\rangle_{z}$. This might explain
the less frequent visits to the zero eigenstates under weak measurement. 

In the strong measurement case illustrated in Figures \ref{fig:zero_eig_avoided}c,d,
the system has ample time to complete a two-step collapse and therefore
the eigenstates are explored in accordance with the Born rule statistics.
The system takes around 0.22 time-units to reach either the $|1\rangle_{z}$
or the $|-1\rangle_{z}$ eigenstates starting from $|0\rangle_{x}$
during an $S_{z}$ measurement (based on 10,000 trajectories arriving
at either of these eigenstates with $a_{max}=1$). On the other hand,
when starting in the superposition $\frac{1}{\sqrt{2}}(|-1\rangle_{z}+|0\rangle_{z}$
on the ellipse in the lower half of Figure \ref{fig:ent_traj_ex}a,
the system takes 0.89 time-units to reach one of $|-1\rangle_{z}$
or $|0\rangle_{z}$ during an $S_{z}$ measurement. For a measurement
interval of limited duration this suggests a greater likelihood for
the system to arrive successfully at the $|-1\rangle_{z}$ and $|1\rangle_{z}$
eigenstates compared with the $|0\rangle_{z}$ eigenstate .

\begin{figure}
\begin{centering}
\includegraphics[width=1\columnwidth]{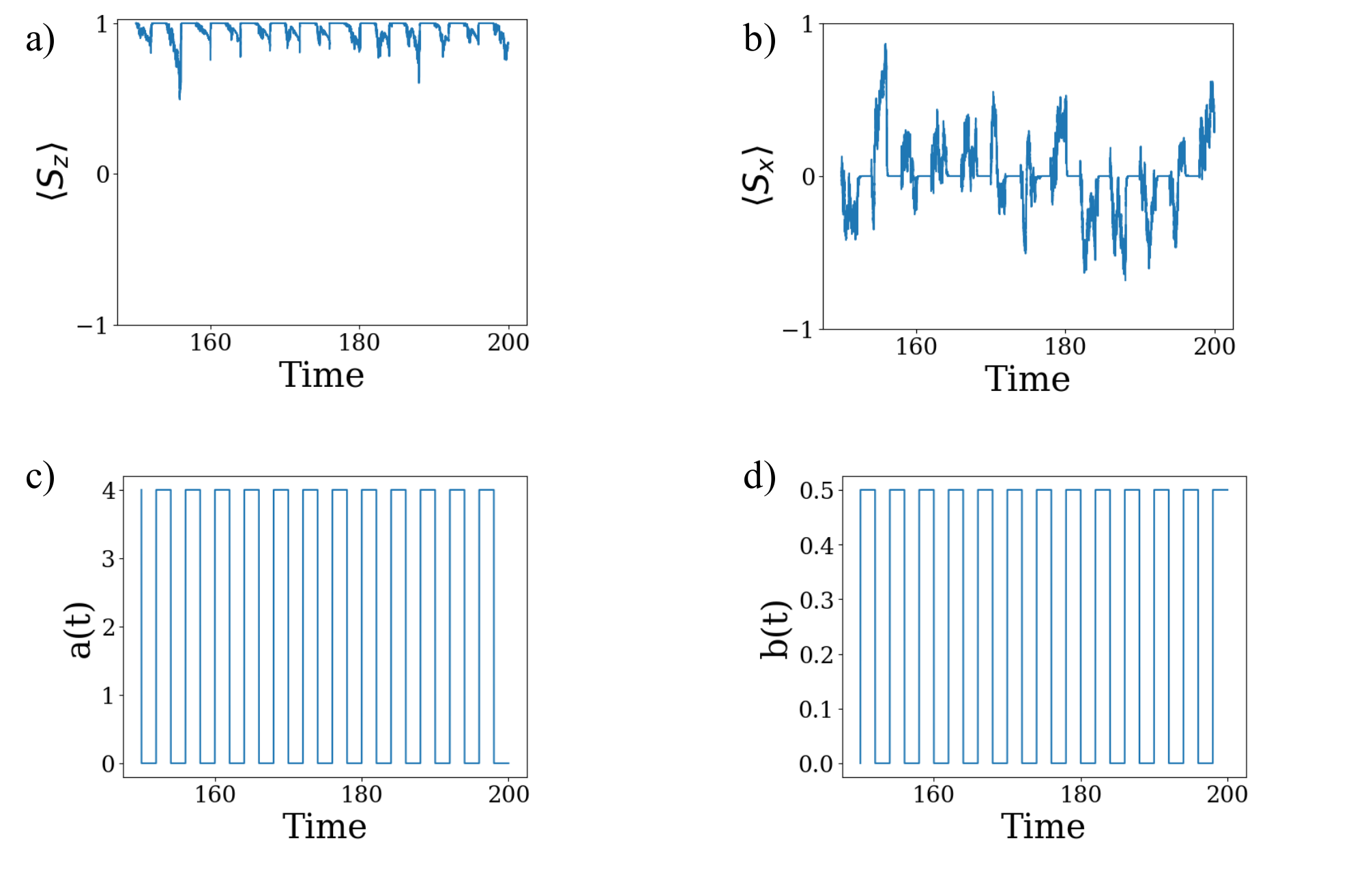}
\par\end{centering}
\caption{a) Evolution of $\langle S_{z}\rangle$: the system returns to the
$|1\rangle_{z}$ eigenstate for more than 13 consecutive $S_{z}$
measurements whilst undergoing a sequence of alternating measurements
of $S_{z}$ and $S_{x}$. b) Evolution of $\langle S_{x}\rangle$.
c) The coupling coefficient $a(t)$ governing the $S_{z}$ measurement.
d) The coupling coefficient $b(t)$ governing the $S_{x}$ measurement.
$S_{z}$ measurement strength $M_{z}=32$; $S_{x}$ measurement strength
$M_{x}=0.5$. Duration 200, time-step 0.0001.\label{fig:spin_1_stab_mech}}
\end{figure}

\begin{figure}
\begin{centering}
\includegraphics[width=1\columnwidth]{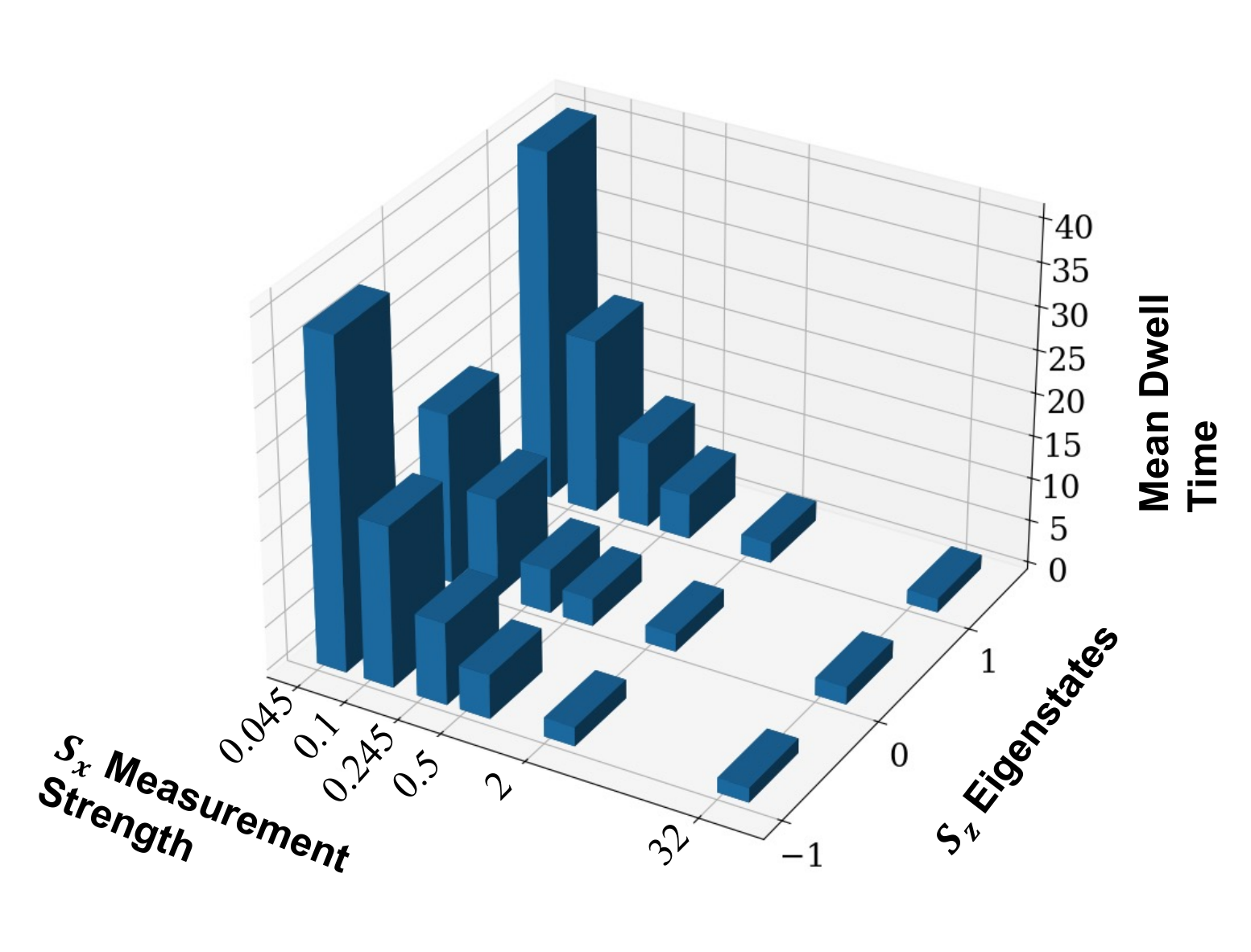}
\par\end{centering}
\caption{The mean dwell time for the three eigenstates of $S_{z}$ at different
$S_{x}$ measurement strengths, for the spin 1 system. $S_{z}$ measurement
strength remains 32 throughout. The dwell time is the number of consecutive
$S_{z}$ measurements for which the system returns to the same eigenstate.
Mean values based on a sequence of 5000 $S_{z}$ measurements. \label{fig:spin_1_dwell_time}}
\end{figure}

We now investigate dwell times for the spin 1 system. Figure \ref{fig:spin_1_stab_mech}
illustrates an example case where a sequence of strong $S_{z}$ measurements
and weak $S_{x}$ measurements produces the return of the system to
the $|1\rangle_{z}$ eigenstate on more than 13 consecutive occasions.
This corresponds to a dwell time of over 13. Figure \ref{fig:spin_1_dwell_time}
depicts the mean dwell time for different $S_{x}$ measurement strengths
and for each eigenstate of $S_{z}$. The $S_{z}$ measurement strength
remains 32 for all cases.

Figure \ref{fig:spin_1_dwell_time} shows how the mean dwell time
at each of the eigenstates of $S_{z}$ increases as the $S_{x}$ measurement
strength is decreased. The mean dwell time appears to be the same
for all three eigenstates of $S_{z}$ at $S_{x}$ measurement strengths
32 and 2, but for lower $S_{x}$ measurement strengths the mean dwell
time is higher for the $|\pm1\rangle_{z}$ eigenstates than for the
$|0\rangle_{z}$ eigenstate. This is a reflection of the decreased
cumulative probability around the $|0\rangle_{z}$ eigenstate depicted
in Figures \ref{fig:spin_1_density_plots}b,d,e,f for low measurement
strengths. 

\section{Conclusions \label{sec:Conclusion}}

We have explored how memory of previous measurement outcomes may survive
in a sequence of alternating measurements of non-commuting observables
for a spin 1/2 system and a spin 1 system. We generate continuous,
diffusive, stochastic quantum trajectories of the reduced density
matrix $\rho$ using the quantum state diffusion (QSD) formalism,
treating measurement as a dynamical interaction between a system and
an environment that is designed to act as a measurement apparatus.
In order to model the sequence of measurements of observables $S_{x}$
and $S_{z}$, we consider the system to interact with two environments
with coupling coefficients $a(t)$ and $b(t)$ modelled as box functions
in time, with $a(t)=1,b(t)=0$ and vice-versa. For the spin 1/2 system
we generate trajectories in terms of the $r_{x},r_{z}$ components
of the coherence vector representing the reduced density matrix. For
the spin 1 system we illustrate trajectories in terms of the spin
components $\langle S_{x}\rangle,\langle S_{z}\rangle$, where $\langle S_{i}\rangle=Tr(\rho S_{i})$.

The spin 1/2 system is drawn towards the eigenstates of $\sigma_{x}$
and $\sigma_{z}$ that lie on the circumference of a Bloch sphere
circle (Figure \ref{fig:rx_rz_example_traj}). The system typically
travels between the eigenstates by remaining on the circle's circumference.
The spin 1 system trajectories may also be represented on a circular
space, though due to the greater number of eigenstates of the observables
in question, the pattern is more complex, with the $|0\rangle_{x}$
and $|0\rangle_{z}$ eigenstates lying at the centre of the circle
and the $|\pm1\rangle_{x}$ and $|\pm1\rangle_{z}$ eigenstates located
on the circumference.  We find that the system avoids certain regions
of the $(S_{x},S_{z})$ space, forming a petal-like pattern of cumulative
probability. 

This is an important feature of the measurement process of the spin
1 system, and we believe that it arises from a multistep process of
collapse, a cascade of rejections of possible measurement outcomes
until just one remains. The system typically starts a measurement
process in a state with non-zero probability amplitudes for all three
eigenstates of the observable in question. The probability amplitude
of one of the eigenstates goes to zero, illustrated in Figure \ref{fig:mes_cascade},
drawing the system towards subspaces representing superpositions of
pairs of eigenstates. The boundaries of the petal-like regions illustrated
in Figure \ref{fig:ent_traj_ex} correspond to these subspaces: the
dynamics takes the system to these boundaries, but does not pass through
them since to do so would be to undo the partial collapse that has
just been performed. Instead, the subsequent dynamics involves eliminating
one or other of the remaining probability amplitudes, taking the system
to an eigenstate, typically along edges of the petals. 

We illustrate the behaviour for both systems using a cumulative probability
density in Figures \ref{fig:spin_half_prob_density} and \ref{fig:spin_1_density_plots}
which reveals that, as the measurement strength is decreased, the
cumulative probability density deviates from that expected of the
Born rule, which demands a strongly peaked probability density around
the eigenstates of $S_{x}$ and $S_{z}$. In particular, the probability
density may be unevenly concentrated around certain eigenstates and
not others, illustrating the system's tendency to return to the eigenstate
it was in at the preceding measurement, demonstrated in Figures \ref{fig:ex_traj_spin_half}
and \ref{fig:spin_1_stab_mech}. For the spin 1 system the cumulative
probability density around the $|0\rangle_{x}$ and $|0\rangle_{z}$
eigenstates decreases starkly as the measurement strength decreases,
and is shifted towards the circumference of the circle where the other
eigenstates of $S_{x}$ and $S_{z}$ reside.

Hence at low measurement strength there is a breakdown in the statistics
predicted by the Born rule. The likely explanation is that an $S_{z}$
measurement is relatively fast to complete when the system starts
in $|0\rangle_{x}$, for which selection of $|\pm1\rangle_{z}$ is
favoured rather than $|0\rangle_{z}$. In contrast, completion of
measurement is typically slower when the system starts in $|\pm1\rangle_{x}$,
which is the initial condition from which the $|0\rangle_{z}$ eigenstate
is potentially reached. This effect arises from the near elimination
of the first step in the multistep collapse process in the former
situation. 

Furthermore, memory of previous eigenstates of $S_{z}$ in the sequence
of alternating measurements of $S_{z}$ and $S_{x}$ can arise from
an incomplete collapse to an eigenstate of $S_{x}$ during a $S_{x}$
measurement. Figures \ref{fig:ex_traj_spin_half} and \ref{fig:spin_1_stab_mech}
demonstrate such a situation, where for several consecutive $S_{z}$
measurements the system is able to return to the same eigenstate of
$S_{z}$ since the system has failed to collapse to an $S_{x}$ eigenstate.
To explore this further we consider sequences of alternating measurements
with strong $S_{z}$ measurements and weak $S_{x}$ measurements.
We calculate the mean dwell time for each eigenstate of $S_{z}$ for
varying $S_{x}$ measurement strength. We define the dwell time as
the number of consecutive $S_{z}$ measurements for which the system
is able to return to the same eigenstate of $S_{z}$.

Figure \ref{fig:spin_1_dwell_time} shows how the mean dwell times
for the spin 1/2 system increase with decreasing $\sigma_{x}$ measurement
strength, with high $\sigma_{z}$ measurement strength throughout.
The mean dwell times appear to be roughly equal for both eigenstates
of $\sigma_{z}$. This is in accordance with the idea that at low
measurement strength the system does not have enough time to collapse
at a $\sigma_{x}$ eigenstate and therefore memory of previous $\sigma_{z}$
outcomes is not completely erased. The system remembers, and can return
to, the same $\sigma_{z}$ eigenstate it was in previously.

For the spin 1 system, the mean dwell time depicted in Figure \ref{fig:spin_1_dwell_time}
is more complicated. The mean dwell time increases with decreasing
$S_{x}$ measurement strength. But whereas at high $S_{x}$ measurement
strengths the mean dwell time is equal for all three eigenstates of
$S_{z}$, at lower $S_{x}$ measurement strengths it is higher for
the $|\pm1\rangle_{z}$ eigenstates than for the $|0\rangle_{z}$
eigenstate. This is reflected in Figures \ref{fig:ex_traj_spin_half}
b-d. We suggest this could be due to the delicate measurement dynamics
of the $|0\rangle_{z}$ eigenstate which lead to it being visited
less frequently than the other two eigenstates.

Sequences of measurements of non-commuting observables have been performed
experimentally with weak measurements \citep{hofmann_2014,suzuki.etal_2016,piacentini.etal_2016},
and we envisage that the investigations outlined in this work could
also be tested experimentally. In addition, wavefunction collapse
has been experimentally observed in real-time, offering insight into
the measurement timescale required to successfully complete a measurement
\citep{murch.etal_2013,jordan_2013}.

In summary, memory of previous outcomes in a sequence of alternating
measurements of non-commuting observables undergoing weak measurements
could arise if there is insufficient time for the system to collapse
to an eigenstate of the second non-commuting observable, introducing
bias towards a return to the eigenstate of the first non-commuting
observable visited previously. This could operate in a similar spirit
to the incomplete measurements exhibited in the partial quantum Zeno
effect \citep{peres.ron_1990,zhang.etal_2019} or reversed measurements,
where the initial quantum state of the system is recovered after a
partial measurement collapse \citep{korotkov.jordan_2006,kim.etal_2012,katz.etal_2008,murch.etal_2013}.
As such, our findings could be useful in fields of quantum state control
such as quantum computation and cybersecurity protocols, quantum state
tomography, as well as quantum error correction protocols where sequences
of non-commuting observables are utilised. The avoidance of the $|0\rangle_{x}$
and $|0\rangle_{z}$ eigenstates of the $S_{x}$ and $S_{z}$ observables
for the spin 1 system could also be of interest in quantum state control.

\section*{Acknowledgements}

SMW is supported by a PhD studentship funded by EPSRC under grant
codes EP/R513143/1 and EP/T517793/1. We thank the referee for their
very insightful and useful comments.

\bibliography{memory_paper_PRA_new_draft}

\appendix

\section{Spin 1 density matrix and SDEs \label{subsec:Appendix-A}}

The Gell-Mann matrices used to form the spin 1 density matrix are
as follows:

\begin{align}
\lambda_{1} & =\begin{pmatrix}0 & 1 & 0\\
1 & 0 & 0\\
0 & 0 & 0
\end{pmatrix}\,\lambda_{2}=\begin{pmatrix}0 & -i & 0\\
i & 0 & 0\\
0 & 0 & 0
\end{pmatrix}\,\lambda_{3}=\begin{pmatrix}1 & 0 & 0\\
0 & -1 & 0\\
0 & 0 & 0
\end{pmatrix}\nonumber \\
\lambda_{4} & =\begin{pmatrix}0 & 0 & 1\\
0 & 0 & 0\\
1 & 0 & 0
\end{pmatrix}\,\lambda_{5}=\begin{pmatrix}0 & 0 & -i\\
0 & 0 & 0\\
i & 0 & 0
\end{pmatrix}\,\lambda_{6}=\begin{pmatrix}0 & 0 & 0\\
0 & 0 & 1\\
0 & 1 & 0
\end{pmatrix}\nonumber \\
\lambda_{7} & =\begin{pmatrix}0 & 0 & 0\\
0 & 0 & -i\\
0 & i & 0
\end{pmatrix}\,\lambda_{8}=\frac{1}{\sqrt{3}}\begin{pmatrix}1 & 0 & 0\\
0 & 1 & 0\\
0 & 0 & -2
\end{pmatrix}.\label{eq:Gell_Mann_matrices}
\end{align}
A parametrisation of the density matrix for the spin 1 system then
emerges:

\begin{equation}
\rho=\frac{1}{3}\begin{pmatrix}1+\sqrt{3}u+z & -i\sqrt{3}m+\sqrt{3}s & \sqrt{3}v-i\sqrt{3}k\\
i\sqrt{3}m+\sqrt{3}s & 1-\sqrt{3}u+z & \sqrt{3}x-i\sqrt{3}y\\
\sqrt{3}v+i\sqrt{3}k & \sqrt{3}x+i\sqrt{3}y & 1-2z
\end{pmatrix}.\label{eq:spin1_density_matrix}
\end{equation}
Stochastic quantum trajectories can be produced from the equations
of motion for the eight variables parametrising $\rho$. These are
generated from

\begin{equation}
dR_{i}=\frac{\sqrt{3}}{2}Tr(d\rho\lambda_{i}),\label{eq:spin1_eqns_motion}
\end{equation}
using the properties of the Gell-Mann matrices. The components of
the coherence vector $\boldsymbol{R}$ defined in section \ref{subsec:Spin-1-system}
are denoted $R_{i}$.

The SDEs for the variables parametrising the spin 1 density matrix
are as follows:

\begin{widetext}

\begin{align}
dk= & -2a^{2}kdt-\frac{1}{2}b^{2}kdt+a\left(-\frac{2}{\sqrt{3}}ku-2kz\right)dW_{z}+b\left(-\frac{2\sqrt{2}}{\sqrt{3}}ks-\frac{2\sqrt{2}}{\sqrt{3}}kx+\frac{1}{\sqrt{2}}m+\frac{1}{\sqrt{2}}y\right)dW_{x}\nonumber \\
dm= & -\frac{1}{2}a^{2}mdt+b^{2}\left(\frac{3}{4}y-\frac{5}{4}m\right)dt-a\left(\frac{2}{\sqrt{3}}mu-2mz+zm\right)dW_{z}+b\left(\frac{1}{\sqrt{2}}k-\frac{2\sqrt{2}}{\sqrt{3}}ms-\frac{2\sqrt{2}}{\sqrt{3}}mx\right)dW_{x}\nonumber \\
ds= & -\frac{1}{2}a^{2}sdt+\frac{1}{4}b^{2}\left(x-s\right)dt+a\left(-\frac{2}{\sqrt{3}}su-2sz+s\right)dW_{z}+b\left(-\frac{2\sqrt{2}}{\sqrt{3}}s^{2}-\frac{2\sqrt{2}}{\sqrt{3}}sx+\frac{1}{\sqrt{2}}v+\frac{\sqrt{2}}{\sqrt{3}}z+\frac{\sqrt{2}}{\sqrt{3}}\right)dW_{x}\nonumber \\
du= & b^{2}\left(-\frac{5}{4}u-\frac{3}{4}v+\frac{\sqrt{3}}{4}z\right)dt+a\left(-\frac{2}{\sqrt{3}}u^{2}-2uz+zu+\frac{1}{\sqrt{3}}z+\frac{1}{\sqrt{3}}\right)dW_{z}+b\left(-\frac{2\sqrt{2}}{\sqrt{3}}su-\frac{2\sqrt{2}}{\sqrt{3}}ux-\frac{1}{\sqrt{2}}x\right)dW_{x}\nonumber \\
dv= & -2a^{2}vdt+b^{2}\left(-\frac{1}{2}v-\frac{3}{4}u+\frac{\sqrt{3}}{4}z\right)dt+a\left(-\frac{2}{\sqrt{3}}uv-2vz\right)dW_{z}+b\left(\frac{2\sqrt{2}}{\sqrt{3}}sv+\frac{1}{\sqrt{2}}s-\frac{2\sqrt{2}}{\sqrt{3}}vx+\frac{1}{\sqrt{2}}x\right)dW_{x}\nonumber \\
dx= & -\frac{1}{2}a^{2}xdt+b^{2}\left(\frac{1}{4}s-\frac{1}{4}x\right)dt-a\left(\frac{2}{\sqrt{3}}ux+2xz+x\right)dW_{z}+\frac{\sqrt{2}}{\sqrt{3}}b\left(-2sx-\frac{\sqrt{3}}{2}u+\frac{\sqrt{3}}{2}v-2x^{2}-\frac{1}{2}z+1\right)dW_{x}\nonumber \\
dy= & -\frac{1}{2}a^{2}ydt+b^{2}\left(\frac{3}{4}m-\frac{5}{4}y\right)dt+a\left(-\frac{2}{\sqrt{3}}uy-2yz-y\right)dW_{z}+b\left(\frac{1}{\sqrt{2}}k-\frac{2\sqrt{2}}{\sqrt{3}}sy-\frac{2\sqrt{2}}{\sqrt{3}}xy\right)dW_{x}\nonumber \\
dz= & b^{2}\left(\frac{\sqrt{3}}{4}u+\frac{\sqrt{3}}{4}v-\frac{3}{4}z\right)dt+a\left(-\frac{2}{\sqrt{3}}uz+\frac{1}{\sqrt{3}}u-2z^{2}-z+1\right)dW_{z}+\frac{\sqrt{2}}{\sqrt{3}}b\left(-2sz+s-2xz-\frac{1}{2}x\right)dW_{x}
\end{align}

\end{widetext}
\end{document}